\documentclass[sigconf,nonacm]{acmart}
\usepackage{multirow}

\usepackage{textcomp}
\usepackage{enumitem}
\usepackage{makecell}

\usepackage{subcaption}
\usepackage[ruled,vlined]{algorithm2e}

\usepackage{stfloats}
\usepackage{pifont}

\newcommand{\SysName}{\texttt{ResiHP}}

\newcommand{\gw}[1]{\textcolor{black}{#1}}
\newcommand{\UCsection}[1]{\section{\MakeUppercase{#1}}}
\begin{document}

\title[\SysName{}: Taming LLM Training Failures with Dynamic Hybrid Parallelism]{\SysName{}: Taming LLM Training Failures with Dynamic Hybrid Parallelism}

\author{Tenghui Ma}
\authornote{
Both authors contributed equally to this work. 
}
\email{24110240061@m.fudan.edu.cn}
\affiliation{%
\institution{Fudan University
  \city{}
  \country{}
  }
\institution{Shanghai AI Laboratory}
}

\author{Jihu Guo}
\authornotemark[1]
\email{24110240025@m.fudan.edu.cn}
\affiliation{%
\institution{Fudan University
  \city{}
  \country{}}
\institution{Shanghai AI Laboratory}
}

\author{Wei Gao}
\authornote{Corresponding author.
}
\email{csgaowei@ust.hk}
\affiliation{%
  \institution{Hong Kong University of Science and Technology
    \city{}
  \country{}}
}

\author{Sitian Lu}
\email{sitianlu@sjtu.edu.cn}
\affiliation{%
\institution{Shanghai Jiao Tong University
  \city{}
  \country{}
}
\institution{Shanghai AI Laboratory}
}

\author{Zhisheng Ye}
\email{yezhisheng@pku.edu.cn}
\affiliation{%
  \institution{Independent Researcher
      \city{}
  \country{}}
}

\author{Hanjing Wang}
\email{wanghanjing@pjlab.org.cn}
\affiliation{%
  \institution{Shanghai AI Laboratory
  \city{}
  \country{}}
}

\author{Dahua Lin}
\email{dhlin@ie.cuhk.edu.hk}
\affiliation{%
  \institution{The Chinese University of Hong Kong
      \city{}
  \country{}}
}

\renewcommand{\shortauthors}{}

\begin{abstract}
Hybrid parallelism underpins large-scale LLM training across tens of thousands of GPUs. At such scale, hardware failures on individual devices lead to \textit{performance skew} across devices, diminishing overall training efficiency. Existing resilient systems overlook sequence length variability in datasets and device performance skew under hybrid parallelism. As a result, (1) iteration time fluctuations induced by sequence length variability can trigger spurious fail-slow detections, and (2) failures are mitigated through individual adaptations in hybrid parallelism, leading to unnecessary detection overhead and inefficient resilient training.

To respond, this paper presents \SysName{}, a resilient system that enables robust failure detection and fine-grained adaptation for hybrid parallel training. First, we develop a \texttt{Detector} to accurately identify failures. In particular, it employs a workload-aware execution time predictor that disentangles failures from iteration time fluctuations while remaining lightweight for online detection. Second, we design a \texttt{Scheduler} that dynamically adapts parallelism group sizes, model partitioning, and workload scheduling policies to improve training efficiency under failures. Experiments show that \SysName{} improves training throughput by 1.04--4.39$\times$ compared with state-of-the-art resilient training systems under diverse failure scenarios in a 256-GPU cluster.
\end{abstract}

\maketitle

\UCsection{Introduction}
Training ever large language models (LLMs) imposes unprecedented demands on computational resources~\cite{megatron2,yang2025qwen3technicalreport,dsv3,openai2024gpt4technicalreport,dubey2024llama}. At today's scale, sustaining high throughput requires hybrid parallelism that combines data parallelism (DP)~\cite{DP}, tensor parallelism (TP)~\cite{shoeybi2019megatron}, pipeline parallelism (PP)~\cite{GPipe}, and others~\cite{megatronSP,DeepSpeedU,lepikhin2020gshard}. However, as cluster scale grows, hardware failures become statistically inevitable~\cite{dubey2024llama,OPT}. These failures commonly appear as \textit{fail-stop} failures, where devices abruptly terminate due to catastrophic faults such as GPU HBM errors~\cite{dubey2024llama,lin2025understanding,OPT,workshop2022bloom}, and \textit{fail-slow} failures, where devices remain operational but degrade in performance and act as stragglers~\cite{PERSEUS,FailSlow,Understandfailslow}.

Despite their different manifestations, both fail-stop and fail-slow introduce \textit{device performance skew}, which impairs training efficiency. We define device performance skew as failure-induced heterogeneity in effective compute and/or communication rates across devices\footnote{Device performance skew differs from stragglers~\cite{lin2025understanding}: skew characterizes the underlying compute/communication rate heterogeneity, whereas stragglers are an execution-level symptom that may arise from skew or from non-failure factors such as workload imbalance.}.
Fail-stop failures force devices offline~\cite{OPT, jang2023oobleck, Bamboo}, reducing the number of active devices in a parallel group (e.g., DP, PP, or TP) and thus lowering its effective service rate~\cite{gandhi2024recycle}. Meta reports that fail-stop failures wasted approximately 178{,}000 GPU hours during the training of OPT-175B~\cite{OPT}.
Fail-slow failures reduce the computation and communication rates of devices~\cite{FailSlow},
triggering a cascading slowdown that originates in TP groups, propagates as bubbles across PP stages, and amplifies global synchronization delays at the DP boundary. 
Recent measurements~\cite{wu2025greyhound} show that 59.2\% of large-scale training jobs ($\geq$ 512 GPUs) encounter fail-slow failures, increasing average job completion time by 34.59\%. Overall, fail-stop and fail-slow introduce significant device performance skew that severely impairs training efficiency.

Prior work~\cite{jang2023oobleck,Bamboo,wu2025greyhound,gandhi2024recycle} generally structures fail-stop and fail-slow failure mitigation as a two-stage protocol:
(1) failure detection, followed by
(2) system-level adaptation to failures.
Fail-stop failures can be identified by periodically collecting execution status from devices, where a device is marked as failed if status collection times out~\cite{jang2023oobleck, Bamboo} or reports explicit error signals~\cite{gandhi2024recycle}. Detecting fail-slow failures is challenging due to the absence of explicit failure indicators~\cite{FailSlow,wu2025greyhound}. The state-of-the-art fail-slow detection approach~\cite{wu2025greyhound} relies on variations in iteration time as a proxy signal to identify candidate fail-slow failures, followed by validation to localize and confirm the degraded devices.
Yet, the iteration time correlates not only with device performance but also with workloads.
Real-world datasets often have diverse sequence lengths, as in the open-source \textit{GitHub} dataset.
Even after applying sequence packing~\cite{SeqPacking, megatron2} to equalize input lengths, the computation workloads can still vary across iterations~\cite{FlexSP, DistTrain,WLB-LLM,gao2025rollpacker} due to the quadratic complexity of self-attention with respect to sequence length~\cite{AttnIs}. \gw{
As a result, workload variability across iterations leads to time fluctuations, which render failure detection prone to false fail-slow positives, thereby incurring unnecessary validation overhead.}

Prior resilient systems~\cite{wu2025greyhound,gandhi2024recycle,jang2023oobleck,wu2025adaptra} adapt to failures by tuning individual dimensions of hybrid parallelism, resulting in suboptimal training efficiency.
ReCycle~\cite{gandhi2024recycle} focuses solely on \emph{PP}-level workload migration to tolerate fail-stop failures.
Oobleck~\cite{jang2023oobleck} and Greyhound~\cite{wu2025greyhound} refine workload redistribution across \emph{DP} groups to balance execution time.
Adaptra~\cite{wu2025adaptra} optimizes \emph{PP}-level workload scheduling to alleviate fail-slow effects.
However, individual optimization in hybrid parallelism fails to address device performance skew efficiently, resulting in workload imbalance (\S~\ref{sec:mot_inef_adap}).
Moreover, they conservatively exclude entire TP groups even when only a subset of devices within a TP group suffer from fail-stop failures, causing hardware waste.
\gw{Overall, prior resilient systems~\cite{wu2025greyhound,gandhi2024recycle,jang2023oobleck,wu2025adaptra} fail to jointly adapt hybrid parallelism to device performance skew, resulting in workload imbalance or low resource utilization.}

These gaps motivate accurate failure detection and progressive system-level adaptation in hybrid parallelism. 
\textbf{Accurate failure detection} requires identifying both fail-stop and fail-slow failures in the presence of iteration-time fluctuations caused by sequence length variability, while remaining lightweight to support online per-iteration detection.
\textbf{Fine-grained system-level adaptation} in hybrid parallelism requires progressively adapting along the TP, PP, and DP dimensions to counter the propagation and amplification of failures.
{\textit{(1) TP-dimension challenge.}} Excluding an entire affected TP group results in severe hardware waste, whereas selectively excluding failed devices to salvage healthy ones introduces complex inter-TP-group communication.
{\textit{(2) PP-dimension challenge.}} Failures exacerbate workload imbalance across PP groups, creating extensive bubbles or stalling the entire pipeline, significantly degrading overall training efficiency.
{\textit{(3) DP-dimension challenge.}} Any remaining imbalance manifests as severe delays at global DP synchronization. Balancing replica completion times must be tightly coordinated with TP and PP adaptations.

To address these challenges, we present \SysName{}, a resilient training system that achieves robust failure detection and efficient system-level adaptation. 
\textbf{For failure detection}, we design a lightweight \texttt{Detector} that accurately identifies both fail-stop and fail-slow failures (\S\ref{sec:det}).
To detect fail-stop failures, the \texttt{Detector} employs a lightweight heartbeat mechanism  to periodically collect heartbeat signals from all devices and marks devices as failed upon heartbeat loss~\cite{jang2023oobleck}.
\gw{To detect fail-slow failures, the \texttt{Detector} adopts online time series analysis~\cite{agudelo2020bayesian} on recorded iteration times to identify device performance degradation.
Specifically, \SysName{} employs an execution time predictor to filter out iteration-time fluctuations caused by sequence length variability, thereby avoiding spurious detections and unnecessary validation while enabling highly accurate and efficient failure identification.}
\textbf{For system-level adaptation}, \gw{we design a \texttt{Scheduler} that progressively mitigates the device performance skew introduced by failures across TP, PP, and DP dimensions.}
{\textit{(1) TP dimension:}} The \texttt{Scheduler} reconfigures TP group sizes to preserve healthy devices whenever possible and improve the effective throughput of affected TP groups. Additionally, it eliminates redundant communication between TP groups of varying sizes to improve communication efficiency (\S\ref{sec:dtr}).
{\textit{(2) PP dimension:}} The \texttt{Scheduler} adaptively repartitions the model to balance iteration time among PP groups. Moreover, it reorders workload execution to efficiently overlap communication and computation (\S\ref{sec:repartition}).
{\textit{(3) DP dimension:}} Guided by the TP and PP adaptations, the \texttt{Scheduler} finally schedules micro-batches across DP groups to balance their execution time (\S\ref{sec:wr}).

Overall, we make the following contributions in this paper.\gw{
\begin{itemize}
    \item We present \SysName{}, a novel framework for resilient LLM training that tames failures with dynamic hybrid parallelism and maximizes throughput.
    \item \SysName{} utilizes an execution time predictor to factor out time fluctuations by sequence length variability, thereby improving detection accuracy and efficiency.
    \item \SysName{} effectively restores training resources and throughput by leveraging fine-grained, system-level adaptation in hybrid parallelism to mitigate failure-induced imbalances.
    \item We implement and evaluate \SysName{} with variants of LLaMA 2~\cite{touvron2023llama} and Qwen 2.5~\cite{yang2024qwen2technicalreport} under diverse failure scenarios in a cluster of 256 A100 GPUs. 
    Experimental results show that \SysName{} achieves approximately 99.4\% failure detection accuracy and improves throughput by 1.04--4.39$\times$ over the baselines \cite{jang2023oobleck,gandhi2024recycle,wu2025greyhound,wu2025adaptra}.
\end{itemize}
}
\UCsection{Background and Motivation}

\subsection{Fail-Stop and Fail-Slow Failures}\label{sec:failures_ana}
\begin{table*}[htbp]
\centering
\caption{Summary of root causes for fail-stop and fail-slow failures in distributed training.}
\vspace{-10pt}
\begin{tabular}{l p{11cm} c}
\toprule
\textbf{Category} & \textbf{Root Causes} & \textbf{Reported Impact} \\
\midrule
\multirow{3}{*}{\textbf{Fail-stop}} 
& \textbf{Hardware:} Memory Error (e.g., OOM and ECC errors), Network Error (e.g., RoCE, NVLink, NIC errors), Node Failure, SSD Storage Error. 
& \multirow{3}{*}{\makecell{Wasting 178,000 GPU hours~\cite{OPT}\\ $\sim$10\% of training time wasted~\cite{dubey2024llama}}} \\
& \textbf{Software:} Data race, Buggy error handling, Indefinite blocking, or loops. 
\\
\midrule
\multirow{3}{*}{\textbf{Fail-slow}} 
& \textbf{Hardware:} Memory pressure, Network degradation (e.g., RoCE, NVLink, NIC issues), CPU contention, Power instability, and Thermal interface anomalies. 
& \multirow{3}{*}{\makecell{34.59\% ACT increase~\cite{wu2025greyhound} and up \\ to 45\% GPU underutilization~\cite{lin2025understanding}}} \\
& \textbf{Software:} Data corruption, Buggy internal checker. 
\\
\bottomrule
\end{tabular}
\label{table:failure_taxonomy}
\end{table*}

Under hybrid parallelism, failures on individual devices introduce device performance skew within and across parallel groups due to inherent synchronization~\cite{megatron2,jiang2024megascale}.
In LLM training, hardware failures primarily manifest as two distinct categories: fail-stop and fail-slow.
We categorize fail-stop and fail-slow failure cases based on an analysis of prior studies~\cite{Understandfailslow, dubey2024llama, Characterization,sun2025ft2} as summarized in Table~\ref{table:failure_taxonomy}.

\noindent\textbf{Fail-stop Failures} refer to deterministic events, such as CUDA errors, NVLink failures, or out-of-memory (OOM) errors, that interrupt the hardware execution immediately. 

\noindent\textbf{Fail-slow Failures} refer to gray failures where a hardware unit remains functional but exhibits reduced efficiency. Unlike fail-stop failures, which trigger immediate termination, fail-slow failures are insidious because they allow the hardware to continue running. Fail-slow failures are often induced by factors such as GPU thermal throttling, HBM3 performance degradation, or network jitter.  

\noindent\textbf{Observations.} 
Table~\ref{table:failure_taxonomy} reveals that fail-stop and fail-slow are fundamentally intertwined rather than isolated phenomena. 
For example, memory and network issues appear as fail-slow when they manifest as performance degradation. Yet, they can escalate into fail-stop once error thresholds are exceeded, timeouts are triggered, or components become unavailable~\cite{SuperBench, Understandfailslow, PERSEUS, IASO}. 
\emph{The shared root causes tightly entangle fail-slow and fail-stop into a coupled failure regime.}

\noindent\textbf{Motivation.}
These observations necessitate a training system that efficiently handles device performance skew caused by both fail-slow and fail-stop failures to preserve training efficiency.

\subsection{Iteration-time Fluctuations}\label{sec:fluc}

\begin{figure}
    \centering
    \includegraphics[width=1\linewidth]{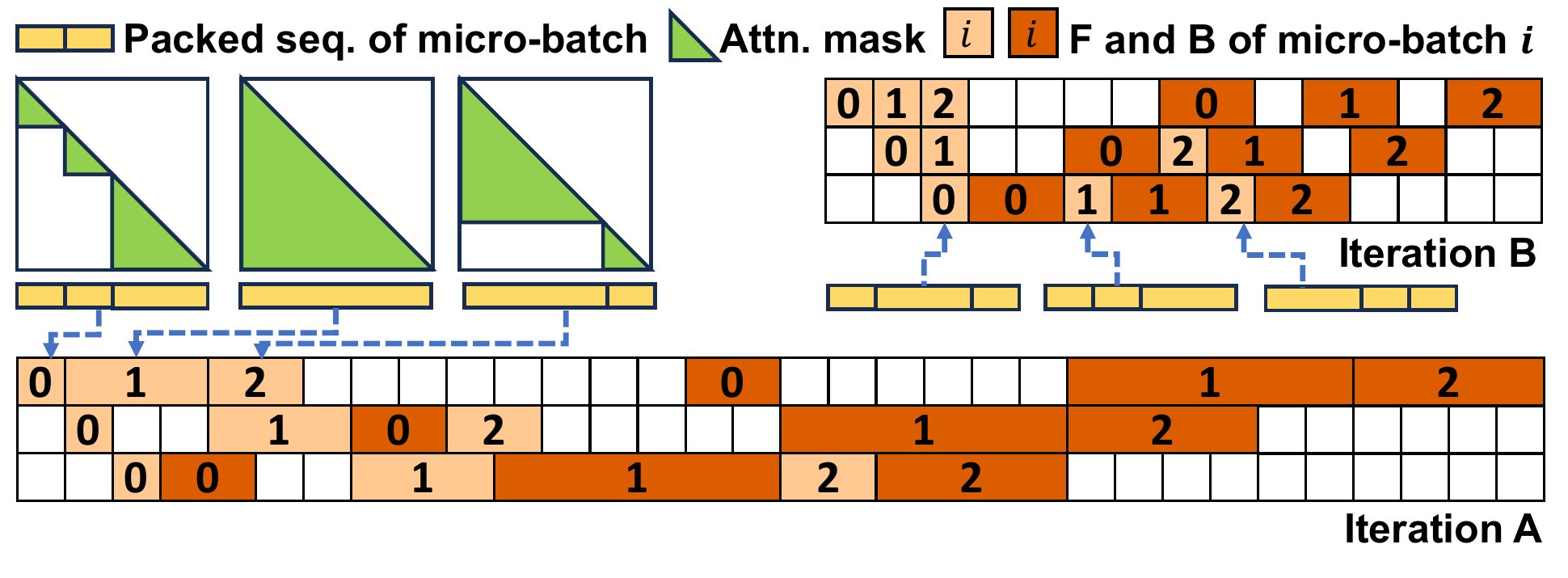}
    \vspace{-20pt}
    \caption{\gw{Illustration of the impact of sequence length variability on iteration-time fluctuations.}}
    \label{fig:fluc}
    \vspace{-15pt}
\end{figure}
\gw{In LLM training, iteration time is inherently confounded by input sequence lengths~\cite{DistTrain,FlexSP,lin2025understanding,Hydraulis}. Figure~\ref{fig:fluc} shows that even with sequence packing~\cite{shoeybi2019megatron,SeqPacking,FlexSP,WLB-LLM}, the quadratic cost of self-attention~\cite{AttnIs} causes substantial computation variability across micro-batches. For example, the attention computation cost of one contiguous 4K-token sequence is about four times that of a packed input of four independent 1K-token sequences. Such variability alters micro-batch execution time, disrupts tightly aligned pipeline schedules, and creates pipeline bubbles due to inter-stage dependencies. Ultimately, it appears as iteration-time fluctuations, which can be misinterpreted as the device performance skew.
}

\noindent\textbf{Motivation.}
\gw{Robust failure detection must account for workload variations to avoid misinterpreting iteration-time fluctuations as device performance skew.}

\subsection{Failure Amplification Effect}

\gw{In hybrid-parallel training, failures usually first manifest within TP. Because TP ranks synchronize frequently within each layer, a single crashed or slow rank can immediately disable or delay its entire TP group. If left unmitigated, this disruption propagates to PP as a degraded or unavailable pipeline stage, and eventually stalls peer DP replicas at global synchronization, amplifying the degradation across the job.}
\gw{To quantify this amplification, we inject a fail-slow failure that halves the speed of one GPU while training LLaMA 2-13B with $(TP, DP, PP)=(4,2,4)$. We measure the number of additionally affected devices and the additional idle GPU time. As shown in Figure~\ref{fig:amplification} (left), one degraded GPU delays 3 additional GPUs in its local TP group, 12 more across the pipeline, and the remaining 16 GPUs at DP synchronization. Figure~\ref{fig:amplification} (right) further shows that additional idle GPU time increases by 4.75$\times$ at TP, 19.13$\times$ at PP, and 25.43$\times$ at DP, relative to the slowdown duration of the faulty GPU.
These results show that a localized failure is substantially amplified as it propagates through the hybrid-parallel hierarchy, eventually affecting the entire 32-GPU job.
}

\noindent\textbf{Motivation.}
\gw{Effective failure mitigation should intervene as early as possible along the failure propagation path, before the effects spread from TP to PP and DP.}

\UCsection{Limitations of existing solutions}
Existing solutions fall short in two respects in improving training efficiency under failures. First, sequence length variability interferes with failure detection. Second, they lack progressive adaptation in hybrid parallelism.

\subsection{High-Overhead Detection}~\label{mot:detect}
Detecting fail-slow failures requires inferring anomalies from indirect signals such as iteration time~\cite{wu2025greyhound}. Although existing detectors can accurately identify iteration-time anomalies, they often fail to distinguish workload-induced fluctuations from device performance skew, especially in long-sequence training, leading to false positives and unnecessary validation overhead.

\subsection{Inefficient System Adaptation}~\label{sec:mot_inef_adap}
\begin{figure}
    \centering
    \includegraphics[width=1\linewidth]{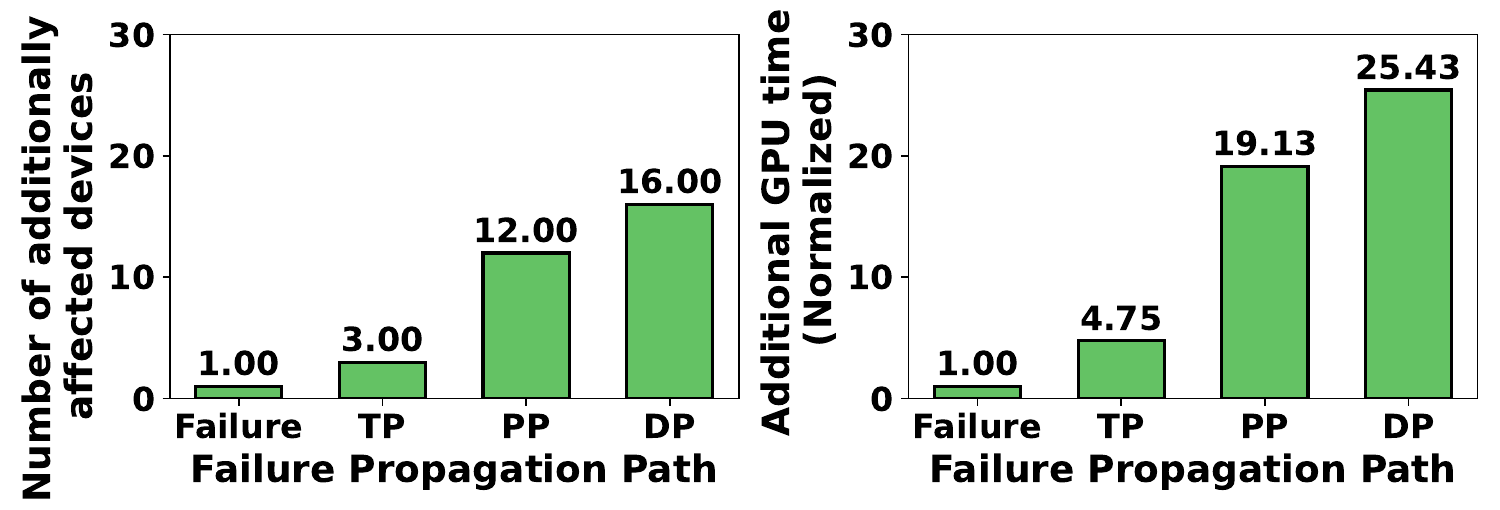}
    \vspace{-20pt}
    \caption{\gw{Failure amplification across TP, PP, and DP under a fail-slow injection on LLaMA 2-13B with $(TP,DP,PP)=(4,2,4)$. 
    }}
    \label{fig:amplification}
    \vspace{-15pt}
\end{figure}

Prior resilient systems typically optimize individual dimensions in hybrid parallelism~\cite{gandhi2024recycle,wu2025greyhound,jang2023oobleck,wu2025adaptra}. Lacking a progressive adaptation mechanism that coordinates across TP, PP, and DP, they suffer from the following critical limitations:

\noindent\textbf{Resource Wastage within TP Groups.}
When a fail-stop failure occurs within a TP group, prior resilient systems~\cite{gandhi2024recycle,jang2023oobleck,wu2025greyhound,wu2025adaptra} conservatively exclude the entire group, even if only a single device has failed. As a result, healthy devices are unnecessarily discarded, leading to severe resource wastage and underutilization.

\noindent\textbf{Inter-DP Imbalance after Workload Migration.}
As shown in Figure~\ref{fig:imbalance}(a), ReCycle~\cite{gandhi2024recycle} tolerates fail-stop failures by migrating workloads at the PP level. When some device in DP0 fails, ReCycle transfers its workloads to DP1, which preserves training progress but introduces significant workload imbalance across DP groups.

\noindent\textbf{Intra-DP Imbalance after Workload Redistribution.}
As shown in Figure~\ref{fig:imbalance}(b), Greyhound~\cite{wu2025greyhound} mitigates inter-DP imbalance by redistributing workloads across DP groups. When DP0 suffers fail-slow failures, its workloads take longer to execute than those in DP1. Greyhound therefore reduces the batch size assigned to DP0 and offloads the remaining samples to DP1 to balance iteration time across DP groups. However, this redistribution introduces workload imbalance among PP groups within a DP group, as illustrated by PP0 and PP1 in DP0. This limitation indicates that effective adaptation requires cross-dimensional coordination to align workloads with device performance skew and improve resource utilization.

{\UCsection{Overview}
\gw{Figure~\ref{fig:overview} presents the overall architecture of \SysName{}. It primarily consists of two key components: the \texttt{Scheduler} and the \texttt{Detector}. The \texttt{Scheduler} orchestrates the distributed training job, dictates progressive system adaptations, and implements the hybrid-parallel execution plan. Meanwhile, the \texttt{Detector} continuously performs lightweight and accurate online failure diagnosis across the cluster. }

\noindent\textbf{Job Launch.}
\gw{Upon job submission (\textcircled{1}), the \texttt{Scheduler} ingests the training configuration to generate an initial execution plan that determines the optimal hybrid-parallel setup and initial workload placement. It then provisions the required computing resources from the GPU pool and dispatches the execution plan (\textcircled{2}).}

\noindent\textbf{Online Monitoring.}
\gw{During training, workers in the GPU pool continuously stream heartbeats and runtime profiling results to the \texttt{Detector} (\textcircled{3}). The \texttt{Detector} analyzes these runtime signals to accurately identify failures (\S\ref{sec:det}). Confirmed failures are summarized into failure reports and promptly sent to the \texttt{Scheduler} (\textcircled{4}).}

\noindent\textbf{System-level Adaptation.}
\gw{Upon receiving a failure report, the \texttt{Scheduler} generates a progressive adaptation strategy based on the current cluster topology and surviving system resources (\S\ref{sec:sche}). It seamlessly reconfigures the parallelism dimensions and redistributes workloads across the active GPU pool, allowing the training process to resume with high efficiency.}

\gw{Overall, this decoupled design preserves training semantics while hiding complex failure mitigation within the system backend.}
\UCsection{Failure Diagnosis} 
\label{sec:det}

\begin{figure}[t]
    \centering
    \includegraphics[width=1\linewidth]{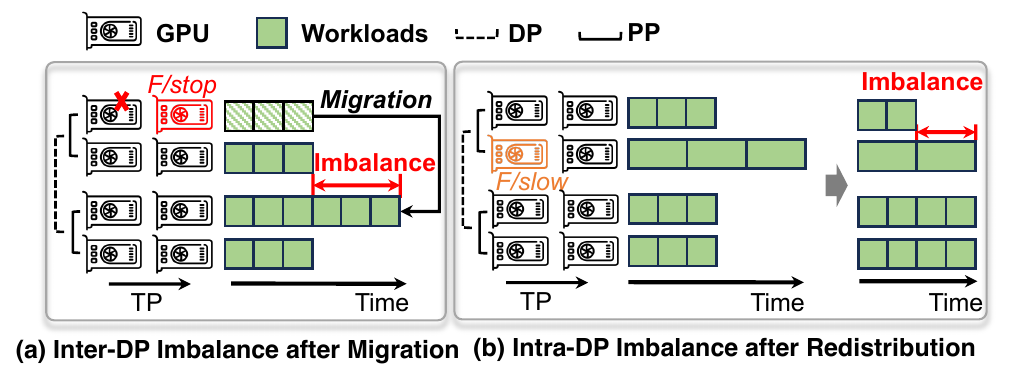}
    \vspace{-20pt}
    \caption{Adapting individual dimensions in hybrid parallelism leads to severe workload imbalance and resource wastage under failures.}
    \label{fig:imbalance}
    \vspace{-10pt}
\end{figure}

This section describes how \texttt{Detector} identifies fail-stop and fail-slow failures during training.

\subsection{Heartbeat-based fail-stop detection}
\gw{To detect fail-stop failures, \texttt{Detector} employs a lightweight, hierarchical two-level heartbeat mechanism.} At the intra-node level, each worker periodically reports compact liveness signals paired with their local training progress. A dedicated node-local monitor aggregates these signals to maintain the active device set and trigger a fail-stop decision upon the absence of several consecutive heartbeats. At the inter-node level, a central coordinator exclusively tracks the status of these node-local monitors, centrally aggregating their fail-stop decisions. \gw{By localizing the raw heartbeat stream and centralizing only the failure decisions, the global monitoring overhead scales gracefully with the number of nodes rather than individual devices, significantly reducing the overhead and complexity of communication across large clusters.}

\subsection{Workload-Aware Fail-Slow Detection}

\begin{figure}[t]
    \centering
    \includegraphics[width=1\linewidth]{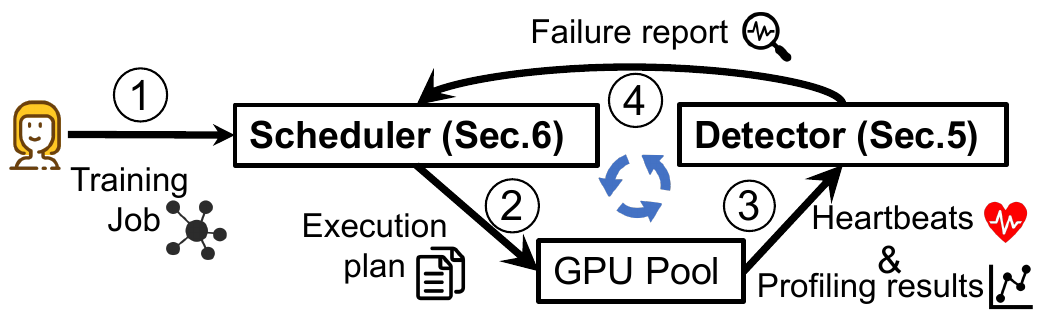}
    \vspace{-20pt}
    \caption{\gw{\SysName{} Overview.}}
    \label{fig:overview}
    \vspace{-15pt}
\end{figure}

\label{sec:workload-aware}
 \gw{To detect fail-slow failures, we analyze the iteration-time series to identify change points and invoke the validation phase to confirm the detection result, following the approach of Greyhound~\cite{wu2025greyhound}. However, iteration-time fluctuations caused by workload variations can trigger false alarms, incurring unnecessary overhead and perturbing the time series for subsequent detections. To avoid this overhead and interference, once a change point is identified, we analytically estimate the expected healthy iteration time under the current workload and pipeline configuration.}

\noindent\textbf{Micro-Batch Time Prediction.}
\gw{We first model the execution time of each micro-batch by separating its linear and quadratic computation components. 
A standard Transformer layer consists mainly of MLP operations and self-attention. For a packed micro-batch with a fixed token budget $N$, MLP computation scales linearly with $N$; since $N$ is fixed across micro-batches, the MLP execution time remains relatively stable. 
In contrast, self-attention has quadratic complexity. With sequence packing, multiple documents with lengths $\{l_1, l_2, \dots, l_k\}$, where $\sum_{i=1}^k l_i = N$, are concatenated with block-diagonal attention masks to prevent cross-document attention. Therefore, the attention cost is proportional not to $N^2$, but to $\sum_{i=1}^k l_i^2$. Based on this property, we model the expected micro-batch execution time as
\begin{equation}
T_{\mathrm{MB}} \approx \alpha N + \beta \sum_{i=1}^k l_i^2,
\end{equation}
where $\alpha$ and $\beta$ capture hardware- and model-specific costs profiled during an initial warm-up phase.}

\noindent\textbf{Iteration-time Prediction.}
\gw{To derive the expected healthy iteration time, \texttt{Detector} uses a lightweight DAG-based analytical simulator that follows the exact pipeline schedule. For each micro-batch $m$ at pipeline stage $s$, the computation is decomposed into $F_{m,s}$, $B_{m,s}$, and $W_{m,s}$, which denote the Forward (F), Backward-Activation (B), and Backward-Weight (W) chunks, respectively~\cite{narayanan2019pipedream,qi2024zero}. The execution time of each chunk is estimated using the micro-batch time predictor.}
\gw{We formulate the pipeline schedule as a DAG $\mathcal{G} = (\mathcal{V}, \mathcal{E})$. Each vertex $v \in \mathcal{V}$ represents a computation chunk, such as $F_{m,s}$, with execution cost $T_{\mathrm{cost}}(v)$. The directed edges in $\mathcal{E}$ encode two scheduling constraints. First, \textit{data dependencies} ensure that a chunk can start only after its input tensor or gradient becomes available. For example, $F_{m,s}$ depends on $F_{m,s-1}$, and the corresponding edge carries the point-to-point communication time $T_{\mathrm{P2P}}$. Similar edges model gradient dependencies during the backward pass. Second, \textit{resource dependencies} encode that each device executes only one chunk at a time and follows the order specified by the pipeline schedule. Thus, consecutive chunks assigned to the same stage are connected by zero-weight edges, ensuring that the next chunk starts only after the previous one completes.}

\gw{Given this DAG, \texttt{Detector} computes the earliest start time of each vertex by topological traversal:
\begin{equation}
t_{\mathrm{start}}(v) =
\max_{u \in \mathrm{pred}(v)}
\left(
t_{\mathrm{start}}(u) + T_{\mathrm{cost}}(u) + T_{\mathrm{edge}}(u, v)
\right),
\end{equation}
where $\mathrm{pred}(v)$ denotes the set of predecessor chunks that must complete before $v$ can start, $t_{\mathrm{start}}(u)$ is the start time of predecessor $u$, $T_{\mathrm{cost}}(u)$ is its execution time, and $T_{\mathrm{edge}}(u,v)$ is the edge cost from $u$ to $v$, such as P2P communication time for data dependencies and zero for resource-ordering edges. This recurrence states that a chunk can start only after all its predecessors have completed and any required communication has finished; hence its start time is determined by the latest satisfied dependency. The expected healthy iteration time is then the completion time of the final sink vertex, i.e., the critical-path length of the DAG.}

\gw{Finally, if the observed iteration time of a change point exceeds the predicted healthy time by more than 25\%, \texttt{Detector} triggers the subsequent validation phase. Otherwise, it treats the iteration as benign, removes the point from the time series, and skips validation.}


\begin{figure}[t]
    \centering
    \includegraphics[width=1\linewidth]{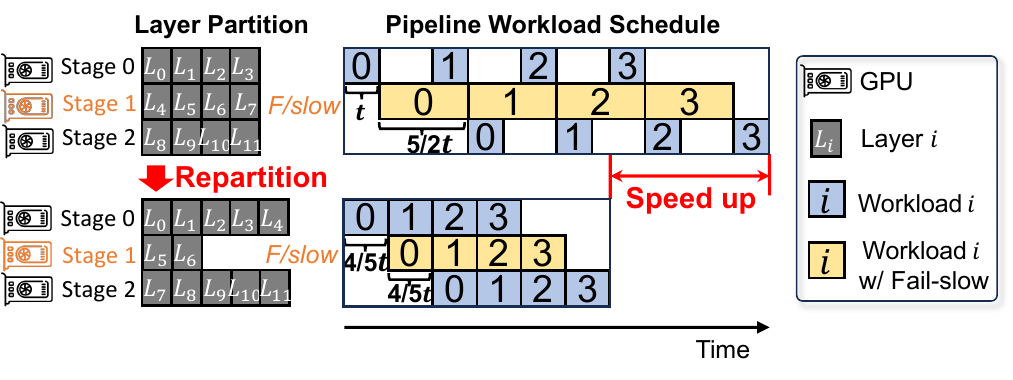}
    \vspace{-20pt}
    \caption{Alleviating PP imbalance via layer repartition.}
    \label{fig:ada_pp}
    \vspace{-10pt}
\end{figure}

\UCsection{Hybrid-parallel Scheduling}\label{sec:sche}
To mitigate the amplified impact of failures, the \texttt{Scheduler} performs progressive adaptations across the TP, PP, and DP.

\subsection{\gw{Selective Device Exclusion Within Affected TP Groups}}
\label{sec:dtr}

When failures occur within a TP group, \texttt{Scheduler} dynamically reconfigures the affected group by selectively excluding failed or severely degraded devices, rather than conservatively removing the entire TP group. The goal is to preserve training continuity while maximizing the effective throughput of the reconfigured TP group.

\noindent\textbf{Candidate TP-Degree Generation.}
Let $G$ denote the original TP group, let $F_{\mathrm{stop}}$ be the set of fail-stop devices in $G$, and let $G' = G \setminus F_{\mathrm{stop}}$ be the remaining executable device set.
\texttt{Scheduler} first determines the feasible range of candidate TP degrees. The maximum executable TP degree is bounded by $|G'|$. The lower bound $k_{\min}$ is dictated by device memory limits, i.e., the minimum TP degree required for each device to hold its model shards. \gw{To preserve attention-head divisibility, communicator layout compatibility, and efficient kernel support~\cite{shoeybi2019megatron}, candidate TP degrees are restricted to powers of two.} Thus, the feasible TP-degree set is:
\begin{equation}
\mathcal{K} =
\{ k \mid k_{\min} \le k \le |G'|,\; k = 2^q,\; q \in \mathbb{Z}_{\ge 0} \}.
\end{equation}

\noindent\textbf{Throughput-Aware Subgroup Selection.}
\gw{For each feasible degree $k \in \mathcal{K}$, \texttt{Scheduler} constructs a candidate subgroup $S_k \subseteq G'$ by greedily selecting the top-$k$ devices ranked by normalized throughput $p_i$, where $p_i$ is measured relative to the healthy peak of device $i$.} This strategy naturally prioritizes healthy devices and includes the fastest fail-slow devices when necessary.
After generating all candidate subgroups, \texttt{Scheduler} selects the optimal one. Because TP relies on tightly synchronized collectives, the effective speed of a TP group is bottlenecked by its slowest member. Meanwhile, a larger TP degree increases aggregate compute throughput by distributing computation across more devices. \gw{To balance parallel scale-out against straggler penalties, \texttt{Scheduler} selects the subgroup $S^*$ that maximizes the estimated aggregate throughput}:
\begin{equation}
S^* = \arg\max_{S_k,\; k \in \mathcal{K}}
\left( k \cdot \min_{i \in S_k} p_i \right).
\label{equ:eff_perf}
\end{equation}

\gw{Due to the power-of-two constraint, the selected subgroup may still leave some healthy or moderately degraded devices unassigned. \texttt{Scheduler} keeps these devices online as node-local standby devices, allowing the system to reuse them for subsequent intra-node failures.} 
Furthermore, the adaptation results in heterogeneous TP degrees, which severely complicates the point-to-point (P2P) communication between different TP groups. To address this, we design a symmetric mapping rule to efficiently orchestrate inter-TP-group communication, the details of which are deferred to the P2P communication optimization in \S\ref{sec:exe}.

In summary, selective exclusion isolates severe failures and salvages partial TP computation capacity. However, the remaining throughput heterogeneity across TP groups can propagate to PP, where it turns the affected stage into a straggler and necessitates subsequent PP-level compensation.

\subsection{Layer Repartition to Alleviate PP Imbalance}
\label{sec:repartition}
To alleviate the straggler introduced after TP adaptation, \texttt{Scheduler} assigns fewer layers to the straggling PP stages and evenly redistributes the excess layers across the remaining stages.
\texttt{Scheduler} adjusts the number of layers assigned to each device to alleviate workload imbalance caused by fail-slow failures.
Figure~\ref{fig:ada_pp} illustrates a pipeline where a fail-slow failure on Stage 1 increases the per-workload computation time. This slowdown propagates to Stage 0 and Stage 2 as pipeline bubbles, forcing the execution time of all PP groups to synchronize with the degraded PP group.
To reduce this imbalance, \texttt{Scheduler} repartitions layers across PP groups. Specifically, it reduces the number of layers on the PP group with fail-slow failures and reallocates them to healthy PP groups.
Figure~\ref{fig:ada_pp} shows the resulting layer repartition, changing the number of layers per PP group from $(4, 4, 4)$ to $(5, 2, 5)$.
This repartition mitigates execution time imbalance.

\subsection{DP Adaptation via Progress-Aware Workload Migration}
\label{sec:wr}
\gw{After TP and PP adaptations, residual execution skew may still remain across DP groups due to heterogeneous TP configurations or unresolved pipeline bubbles. To better align DP completion times, \texttt{Scheduler} dynamically migrates micro-batch workloads across DP groups at the granularity of individual PP stages. We formulate this fine-grained migration as a constrained makespan minimization problem and solve it using an online progress-aware heuristic.}

\noindent\textbf{Problem Formulation.}
\gw{A micro-batch normally executes all its stages within its source DP group. To tolerate failures or mitigate stragglers, the \texttt{Scheduler} may migrate the workloads of one stage to the corresponding stage in another DP group. The migration decision is represented by $x_{i,j}^{d\to d'}$. The objective is to minimize the iteration time, i.e., the maximum completion time across DP groups:
\begin{equation}
    \min_{\mathbf{X}} \max_{d \in \mathcal{D}} T_{\mathrm{makespan}}(d)
\end{equation}
subject to dependency and resource constraints.}

\noindent\textbf{Scheduling Constraints.}
\gw{
The migration plan must satisfy three constraints.
(1) \textit{Execution completeness.} Each stage of each micro-batch is executed exactly once:
$\sum_{d' \in \mathcal{D}} x_{i,j}^{d \to d'} = 1$.
(2) \textit{Dependency preservation.} If stage $i$ of a micro-batch from $d$ is migrated to $d'$, activations and gradients must be exchanged between $d$ and $d'$ so that the adjacent stages on the original pipeline can continue execution.
(3) \textit{Memory capacity.} The memory footprint on the destination stage must not exceed its capacity, i.e., $M_{d',i} \le C_{d',i}$ at any time, including live activations that remain until the corresponding backward computation completes.}

\begin{table}[t]
\centering
\setlength{\tabcolsep}{3pt}
\caption{\gw{Notation used in DP workload migration.}}
\label{tab:dp_notation}
\vspace{-10pt}
\begin{tabular}{ll}
\toprule
Symbol & Meaning \\
\midrule
$\mathcal{D}$ & Set of DP groups \\
$\mathcal{S}$ & Set of PP stages, $\mathcal{S}=\{0,1,\dots,I-1\}$ \\
$d,d'$ & Source and executor DP groups \\
$i,j,t$ & PP stage, micro-batch, and scheduling time slot \\
$x_{i,j}^{d\to d'}$ & 1 iff micro-batch $j$ of stage-$i$ from $d$ runs on $d'$ \\
$T_{\mathrm{makespan}}(d)$ & Completion time of DP group $d$ \\
$M_{d',i}, C_{d',i}$ & Memory footprint and capacity of stage $i$ on $d'$ \\
$P_{d,i,t}$ & Progress of stage $i$ in DP group $d$ at time slot $t$ \\
$d_{\min}, d_{\max}$ & Slowest and fastest DP groups for stage $i$ \\
$\delta$ & Progress-imbalance threshold \\
$F_{j,i,d}$ & F chunks of micro-batch $j$ at stage $i$ in $d$ \\
\bottomrule
\end{tabular}
\vspace{-8pt}
\end{table}

\noindent\textbf{Progress-Aware Heuristic Solver.}
\gw{
Solving the global migration problem as a mixed-integer program is too expensive for online training. Instead, the \texttt{Scheduler} uses a progress-aware heuristic with an analytical pipeline simulator, as shown in Algorithm~\ref{alg:dp_adaptation} and Figure~\ref{fig:wm}. The computation of each micro-batch is decomposed into Forward (F), Backward-Activation (B), and Backward-Weight (W) chunks.}
\gw{To quantify execution progress \textcircled{1}, \texttt{Scheduler} maintains $P_{d,i,t}$ for each stage and DP group. Specifically, $P_{d,i,t}$ counts the number of forward workloads completed by stage $i$ in DP group $d$, including both local and migrated micro-batches. At each scheduling iteration, \texttt{Scheduler} advances the execution state and identifies the slowest and fastest DP groups for each stage:
\begin{equation}
    d_{\min} = \arg\min_{d \in \mathcal{D}} P_{d,i,t}, \quad
    d_{\max} = \arg\max_{d \in \mathcal{D}} P_{d,i,t}.
\end{equation}}

\begin{algorithm}[t]
\caption{\gw{Progress-Aware Workload Migration}}
\label{alg:dp_adaptation}
\While{unfinished workloads exist}{
    Advance pipeline schedules and update dependencies\;
    \ForEach{PP stage $i \in \mathcal{S}$}{
        \tcp{\textcircled{1} Identify slow/fast replicas}
        Compute $P_{d,i,t}$ for all $d \in \mathcal{D}$\;
        $d_{\min} \gets \arg\min_d P_{d,i,t}$, \quad
        $d_{\max} \gets \arg\max_d P_{d,i,t}$\;
        \tcp{\textcircled{2} Generate migration plan}
        \If{$(d_{\min},i)$ is fail-stop \textbf{or} $P_{d_{\max},i,t} - P_{d_{\min},i,t} > \delta$}{
            $j \gets \textsc{NextPending}(d_{\min},i)$\;
            \tcp{\textcircled{3} Migrate if memory is feasible}
            \If{$j \neq \bot$ \textbf{and} $\textsc{MemoryFeasible}(j, i, d_{\max})$}{
                Migrate stage-$i$ workload of $j$ to $d_{\max}$\;
                Update $P_{d_{\min},i,t}$ and $P_{d_{\max},i,t}$\;
            }
        }
    }
}
\end{algorithm}

\gw{As shown in Figure~\ref{fig:wm}(b), at time slot $T_2$, the progress metrics for stage 0 are $P_{0,0,2}=P_{2,0,2}=2$ and $P_{1,0,2}=1$. This identifies DP1 as the straggler ($d_{\min}$), while DP0 and DP2 tie for the maximum progress ($d_{\max}$). Guided by the detected failures and measured progress, the \texttt{Scheduler} evaluates the following migration decisions \textcircled{2}:
\begin{itemize}[leftmargin=*]
    \item \textbf{Fail-slow load balancing.}
   The \texttt{Scheduler} triggers workload migration only when the progress gap exceeds a predefined threshold $\delta$ (i.e., $P_{d_{\max},i,t} - P_{d_{\min},i,t} > \delta$). As illustrated in Figure~\ref{fig:wm}(b), at time $T_2$, when $\delta = 0$, the observed progress gap is $2 - 1 = 1 > 0$. Consequently, stage 0 in both DP0 and DP2 are eligible migration destinations for the pending chunk $F_{5,0,1}$ from the degraded DP1.
    \item \textbf{Fail-stop eviction.}
    If one stage encounters a fail-stop failure, it can no longer execute its remaining workloads. The \texttt{Scheduler} therefore migrates its pending micro-batches to healthy peer stages in other DP groups. In Figure~\ref{fig:wm}(a), stage 2 of DP1 crashes, so the pending stage-2 workloads of DP1 are migrated to peer stages in other DP groups. When DP0 and DP2 have the same progress at this stage, both are valid destinations.
\end{itemize}
Each migration is first simulated and finalized only if the destination stage satisfies the memory constraint at the projected arrival time \textcircled{3}. 
By traversing all stages and tracking their progress globally, the \texttt{Scheduler} handles fail-stop and fail-slow failures in a unified manner and performs fine-grained stage-level migration subject to memory feasibility. This global yet lightweight heuristic greedily reduces inter-DP progress gaps and balances DP completion times with negligible scheduling overhead.}

\begin{figure*}[t]
\centering
\includegraphics[width=\textwidth]{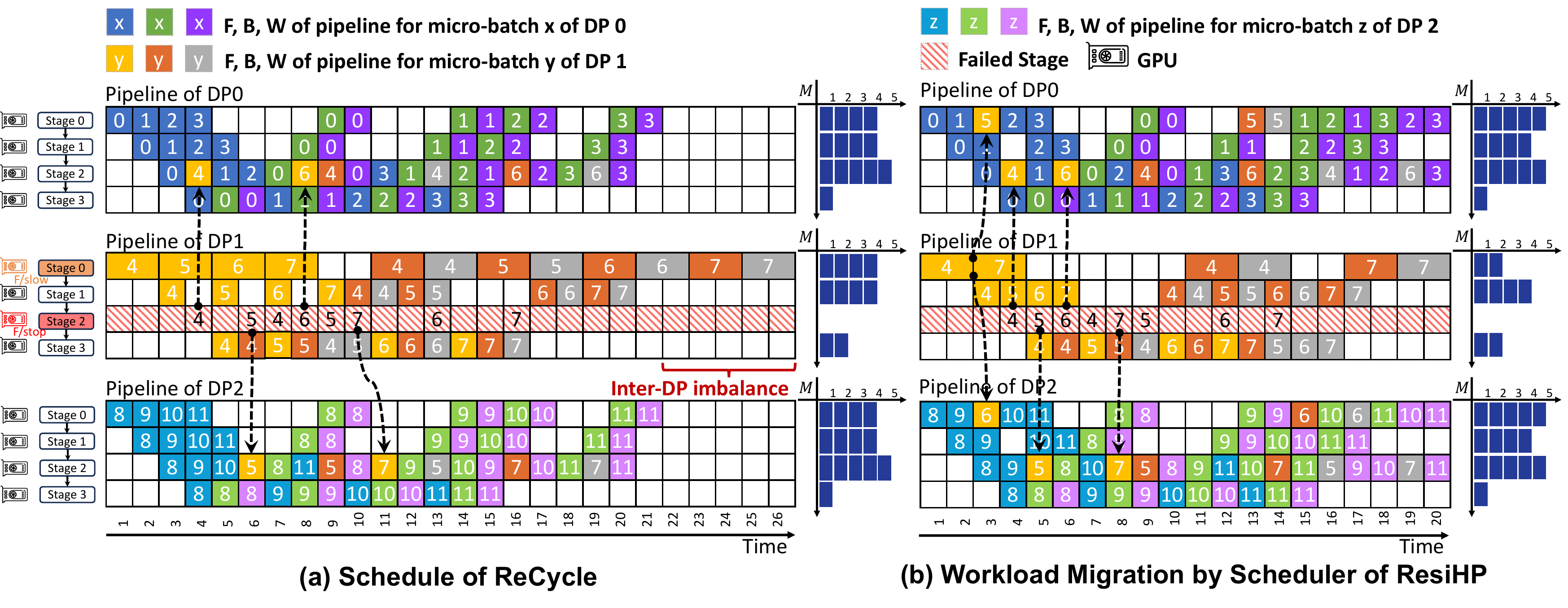}
\vspace{-25pt}
\caption{\gw{(a) ReCycle \cite{gandhi2024recycle} migrates failed-stage workloads without considering stage-level progress, causing inter-DP imbalance. 
(b) \texttt{Scheduler} migrates pending workloads to faster peer stages under memory constraints, thereby reducing imbalance and shortening iteration time.}}
\label{fig:wm}
\end{figure*}

\UCsection{Implementation \& optimization}~\label{sec:exe}
We have implemented \SysName{} in approximately 9k lines of Python code based on our internal optimized LLM training framework, similar to Megatron-LM\cite{megatron2}. 

\noindent\textbf{Detector.}
\texttt{Detector} employs two decoupled runtime mechanisms to identify fail-stop and fail-slow failures. \gw{For immediate fail-stop detection, we use a TCP-based side channel: a CPU agent on each node maintains a persistent TCP connection to a centralized controller. When a node crashes, the controller detects the socket disconnection and broadcasts a fail-stop notification to trigger system reconfiguration.
\texttt{Detector} preserves Greyhound's validation-based fail-slow detection criterion while introducing a workload-aware filtering step prior to validation. }

\begin{figure}[t!]
    \centering
    \includegraphics[width=1\linewidth]{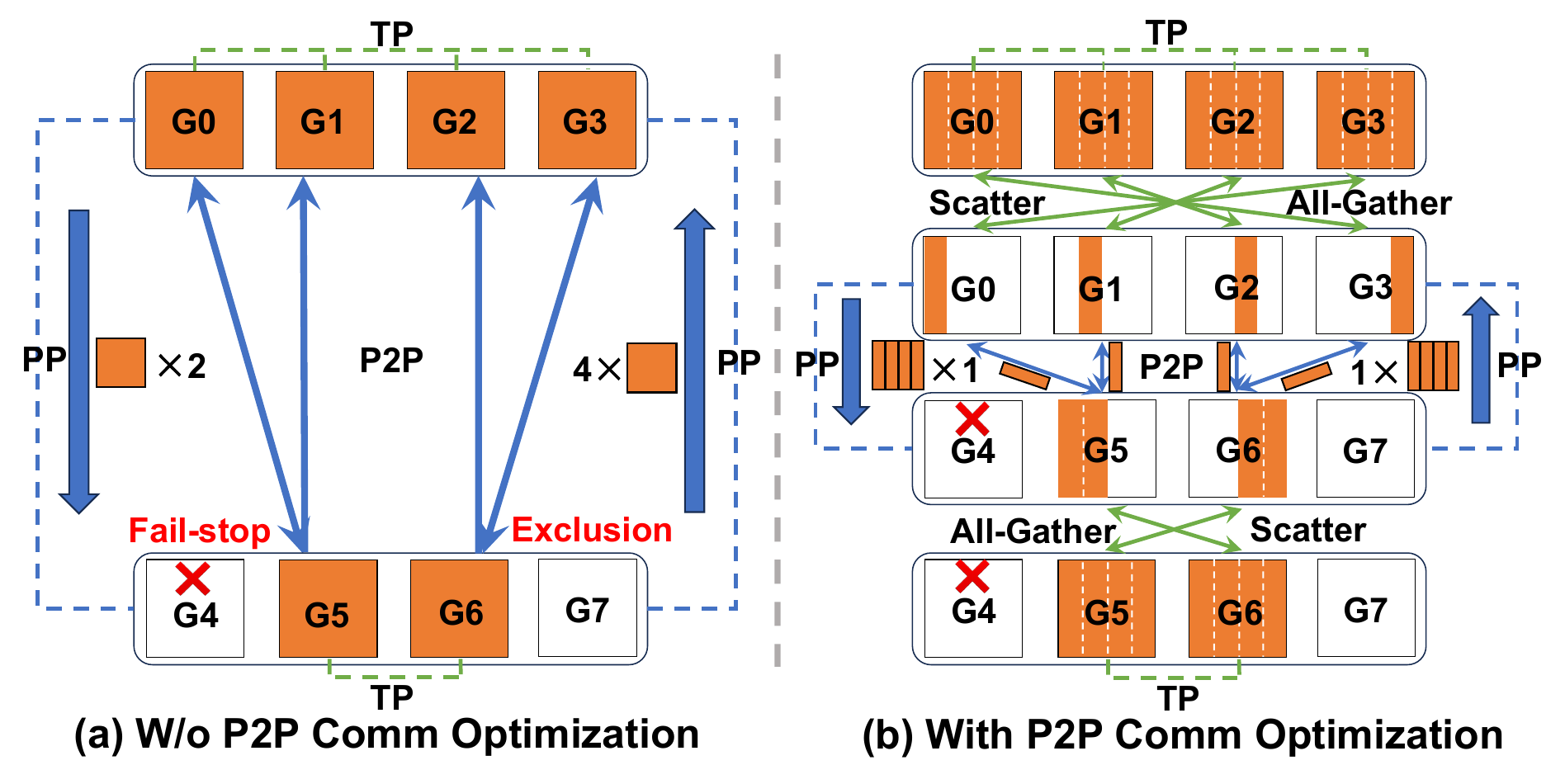}
    \vspace{-20pt}
    \caption{Eliminating redundant P2P transfers after dynamic TP reconfiguration. Left: without scatter/gather, identical tensors are sent repeatedly (2$\times$ top-down and 4$\times$ bottom-up) across nodes over InfiniBand. Right: with scatter/gather, tensors are first scattered or gathered over the faster intra-node NVLink/NVSwitch fabric, and only one copy is sent across nodes over InfiniBand, reducing communication traffic.}
    \label{fig:p2p}
\end{figure}

\noindent\textbf{Scheduler.}
The \texttt{Scheduler} receives failure diagnostics from the \texttt{Detector}, including failure location and severity, and generates an adaptation plan that specifies layer partitioning and workload scheduling. \gw{During reconfiguration, we use \texttt{torch.\allowbreak distributed} to destroy stale communication groups and rebuild them while excluding ranks affected by fail-stop failures; the model is then reshaped across the reconstructed groups according to the new partitioning plan. For runtime execution, \texttt{Scheduler} emits a sequence of primitives, such as \texttt{Forward}, \texttt{Backward}, \texttt{Send}, and \texttt{Recv}, which are parsed and executed by a lightweight worker-side interpreter.}

\noindent\textbf{Portability.}
\gw{The ideas of the \texttt{Detector} and \texttt{Scheduler} are portable to other hybrid-parallel frameworks with analogous control over communication groups and execution schedules, while communicator reconstruction and primitive dispatch are framework-specific.}

\begin{figure}[t]
    \centering
    \includegraphics[width=1\linewidth]{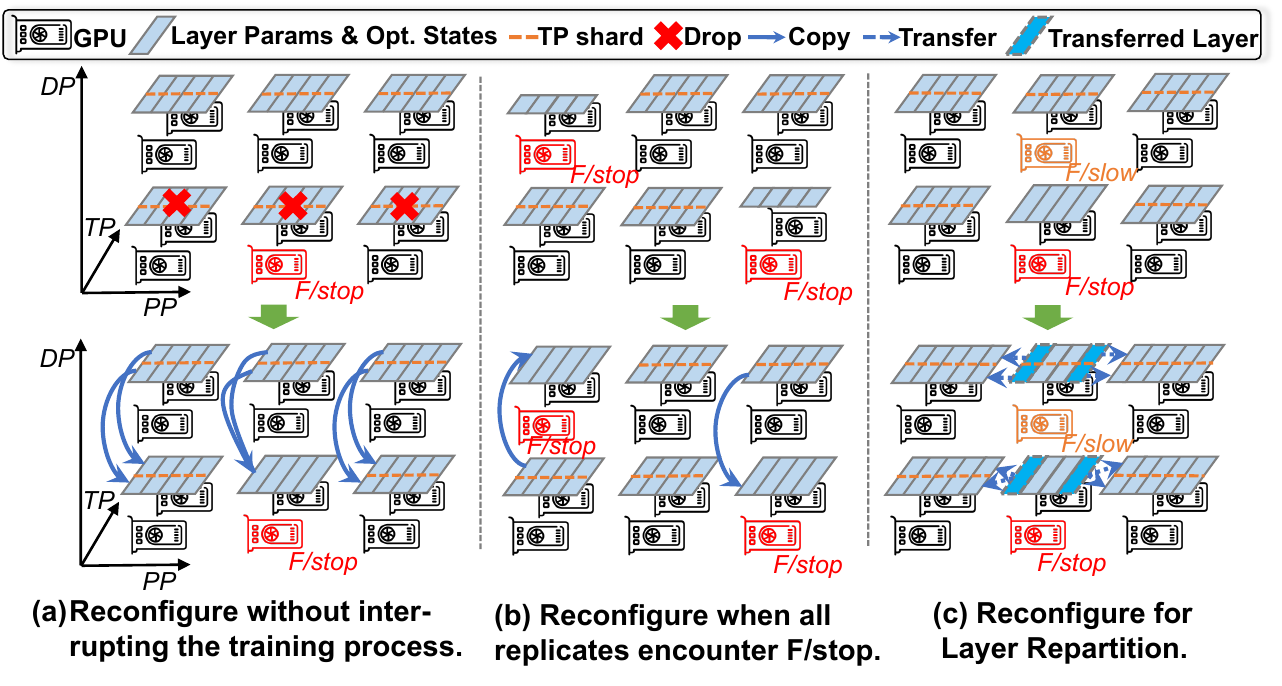}
    \vspace{-20pt}
    \caption{\gw{Recovery of optimizer states and parameters during reconfiguration. ResiHP copies or transfers states from surviving replicas to rebuild dropped TP shards, support layer repartition, and recover training states when all replicas in a stage encounter fail-stop failures.}}
    \label{fig:state}
    \vspace{-15pt}
\end{figure}

\noindent\textbf{P2P Communication Optimization.}
During P2P communication between adjacent pipeline stages, GPUs within the same TP group send and receive identical tensors. As shown in Figure~\ref{fig:p2p}(a), without scatter/gather optimization, the same tensor set is transmitted twice in the top-down direction and four times in the bottom-up direction over InfiniBand, causing substantial cross-node communication redundancy. To reduce this redundancy, we build on Megatron-LM’s scatter/gather optimization~\cite{megatron2}. This optimization requires identical TP degrees for both the sender and receiver in P2P communication. However, dynamic TP reconfiguration (\S\ref{sec:dtr}) for handling fail-stop failures may leave communicating peers with heterogeneous TP degrees, violating this requirement. To preserve scatter/gather in this setting, we introduce new P2P communication rules. As shown in Figure~\ref{fig:p2p}(b), the sender first scatters tensors into $N$ equal-sized chunks, where $N$ is the larger TP degree of the two peers, and sends the chunks to the corresponding GPUs over InfiniBand. The receiver then reconstructs the final tensor via a faster intra-node NVLink/NVSwitch all-gather, so that identical tensors are sent only once across nodes, thereby reducing InfiniBand traffic and improving communication efficiency.

\begin{table}[t!]
  \centering
  \caption{\gw{Models and 3D parallelism settings.}}
  \vspace{-8pt}
  \setlength{\tabcolsep}{4pt}
  \renewcommand{\arraystretch}{1.15}
  \begin{tabular*}{\columnwidth}{@{\extracolsep{\fill}}lcccc@{}}
    \toprule
    \textbf{Scale} & \textbf{LLaMA 2 ~\cite{touvron2023llama}} & \textbf{Qwen 2.5 ~\cite{yang2024qwen2technicalreport}} & \textbf{(TP,DP,PP)} & \textbf{\#GPUs} \\
    \midrule
    Small  & 7B  & 7B  & (4, 2, 2)   & 16 \\
    Medium & 13B & 14B & (4, 2, 4)  & 32 \\
    Large  & 30B & 32B & (4, 2, 8)  & 64 \\
    XLarge & 70B & 72B & (4, 4, 16) & 256 \\
    \bottomrule
    \end{tabular*}
  \vspace{-10pt}
  \label{tab:models}
\end{table}

\begin{table}[t]
  \centering
  \caption{\gw{Prediction accuracy of the micro-batch time predictor (MTP) and iteration-time predictor (ITP), reported as Mean Absolute Percentage Error (MAPE).}}
  \vspace{-8pt}
  \setlength{\tabcolsep}{3pt}
  \renewcommand{\arraystretch}{1.15}
  \begin{tabular*}{\columnwidth}{@{\extracolsep{\fill}}lccccc@{}}
    \toprule
    \textbf{Model} & \textbf{Seq.} & \textbf{MBs} & \textbf{Sched.} & \textbf{MTP} & \textbf{ITP} \\
    \midrule
    Qwen 2.5-7B  & 8K  & 4 & 1F1B\cite{narayanan2019pipedream} & 1.19\% & 2.81\% \\
    Qwen 2.5-14B & 16K & 8 & ZBH\cite{qi2024zero}  & 1.58\% & 5.06\% \\
    LLaMA 2-13B  & 32K & 8 & 1F1B & 1.21\% & 4.89\% \\
    \bottomrule
  \end{tabular*}
  \vspace{-10pt}
  \label{tab:prediction-accuracy}
\end{table}

\noindent\textbf{Optimizer State and Parameter Recovery.}
\gw{After a failure is detected, the runtime enters an online reconfiguration phase to reconstruct communication groups, model states, and optimizer states while preserving training progress. As shown in Figure~\ref{fig:state}(a), \SysName{} first excludes failed DP replicas and rebuilds the communication groups, allowing healthy replicas to continue execution. Once the current iteration completes, the latest committed parameters and optimizer states are synchronized across the reconfigured groups.}
\gw{Figure~\ref{fig:state}(b) illustrates the case in which different DP replicas encounter fail-stop failures. In this case, training must pause, and \SysName{} falls back to the persistent states from the last completed iteration to reconstruct the missing states and parameters. Figure~\ref{fig:state}(c) shows recovery under layer repartition: \SysName{} migrates parameters and optimizer states together with reassigned layers, and if the TP degree also changes, it dynamically reshards the transferred states to match the target TP layout using optimized P2P communication.}
\gw{Through this system-level state remapping, \SysName{} preserves training semantics, maintains training progress, and sustains convergence even under frequent failures.}

\UCsection{Evaluation}
We evaluate the \texttt{Detector} and the \texttt{Scheduler} to answer three questions: 
(1) How accurately does the \texttt{Detector} identify fail-stop and fail-slow failures across various models and parallelism settings (\S\ref{sec:accurate})?
(2) How effective is the \texttt{Scheduler} at mitigating fail-stop and fail-slow failures under different failure scenarios and parallelism settings (\S\ref{sec:effective})?
(3) How well does \SysName{} preserve training efficiency in large-scale real-world failure scenarios (\S\ref{sec:scale})?

\subsection{Experimental Setup}

\noindent\textbf{Testbed Configuration.} We conduct our evaluation on a cluster comprising 32 nodes, each equipped with 8 NVIDIA A100 GPUs connected via NVSwitch. The nodes are interconnected via 200Gbps HDR InfiniBand. We utilize our internal optimized framework to train a set of LLaMA 2 and Qwen 2.5 models of various sizes and parallel strategies. Table \ref{tab:models} details the optimized hybrid parallelism strategies for the allocated set of healthy GPUs. The testbed runs CUDA version 12.2 and NCCL version 2.20.1.

\begin{table}[t]
  \centering
  \caption{\gw{Avg. number of false alarms (FA), overhead for one false alarm, and fail-slow detection accuracy over traces. Results are reported as \SysName{} vs. Greyhound~\cite{wu2025greyhound}.}}
  \vspace{-8pt}
  \setlength{\tabcolsep}{4pt}
  \renewcommand{\arraystretch}{1.15}
  \begin{tabular*}{\columnwidth}{@{\extracolsep{\fill}}lccc@{}}
    \toprule
    \textbf{Model Size, Seq.} & \textbf{FA}  & \textbf{Overhead} & \textbf{Accuracy}  \\
    \midrule
    Small, 8K  & \makecell{0 / 3.7  }  &\makecell{34ms / 2.24s}  &\makecell{1.00 / 1.00} \\
    Medium,16K & \makecell{0.1 /5.2 }  &\makecell{45ms / 3.28s} &\makecell{1.00 / 1.00}\\
    Medium, 32K &  \makecell{0.3 / 8.7 } &\makecell{49ms / 3.72s}  &\makecell{0.98 / 0.98}\\
    \bottomrule
  \end{tabular*}
  \vspace{-10pt}
  \label{tab:false-alarms}
\end{table}

\noindent\textbf{Failure injection.} We evaluate the resilience of our system using deterministically injected fail-stop and fail-slow failures. To emulate fail-stop failures, we manually terminate a subset of workers during training, forcing the system to resume execution with the remaining available devices. To simulate computational fail-slow failures, we employ nvidia-smi to lock the GPU SM frequency. To inject communication fail-slow failures, we initiate side-channel communication jobs that create network bandwidth contention, thereby reducing the available bandwidth on specific network links.

\noindent\textbf{Baselines.}
\gw{We compare \SysName{} with four representative baselines: Greyhound~\cite{wu2025greyhound}, Adaptra~\cite{wu2025adaptra}, ReCycle~\cite{gandhi2024recycle}, and Oobleck~\cite{jang2023oobleck}.} They cover three failure-handling capabilities.

\noindent\textit{Fail-slow detection and mitigation.}
Greyhound detects fail-slow failures from anomalous iteration-time increases and mitigates them by redistributing micro-batches across DP groups according to processing speed. \gw{Adaptra optimizes PP-level workload scheduling to mitigate communication-related fail-slow failures.}

\noindent\textit{Fail-stop tolerance.}
ReCycle tolerates fail-stop failures by rerouting micro-batches from failed ranks to DP peers in the same pipeline stage to execute. \gw{Oobleck recovers by switching to a precomputed pipeline template that uses fewer nodes.}

\noindent\textit{Mixed-failure handling.}
\gw{By integrating Greyhound's fail-slow detection and mitigation, we build strengthened versions of ReCycle and Oobleck that can handle both fail-slow and fail-stop failures.}

\noindent\textbf{Metrics.}
\gw{For detection, we report the Mean Absolute Percentage Error (MAPE) of the micro-batch and iteration-time prediction, as well as detection accuracy and  average false alarms over the entire job. To evaluate the effectiveness of \texttt{Scheduler} in handling failures, we report end-to-end throughput in samples per second (samples/s) across training iterations. Unless otherwise stated, each aggregate result is averaged over multiple independent runs, and error bars denote 95\% confidence intervals computed from per-run means~\cite{jain1991art,georges2007statistically}. For single-run temporal traces, we report the observed trajectory without confidence intervals.}

\subsection{Detector Accuracy}

\label{sec:accurate}
\begin{table*}[!ht]
  \centering
  \caption{\gw{Average Throughput (samples/s) with increasing fail-stop failure frequency (– denotes aborted training).}}
  \vspace{-5pt}
  \setlength{\tabcolsep}{3pt}
  \renewcommand{\arraystretch}{1.2}
  \begin{tabular*}{\textwidth}{@{\extracolsep{\fill}}c|ccc|ccc|ccc|ccc|ccc|ccc@{}}
    \noalign{\hrule height 1pt}
    \textbf{Models} &
      \multicolumn{3}{c|}{\textbf{LLaMA 2-7B }} &
      \multicolumn{3}{c|}{\textbf{LLaMA 2-13B }} &
      \multicolumn{3}{c|}{\textbf{LLaMA 2-30B }} &
      \multicolumn{3}{c|}{\textbf{Qwen 2.5-7B}} &
      \multicolumn{3}{c|}{\textbf{Qwen 2.5-14B}} &
      \multicolumn{3}{c}{\textbf{Qwen 2.5-32B}}
      \\
    \cline{2-19}

    \textbf{Fail-stop Frequency} &
      \textbf{2h} & \textbf{1h} & \textbf{30m} &
      \textbf{2h} & \textbf{1h} & \textbf{30m} &
      \textbf{2h} & \textbf{1h} & \textbf{30m} &
      \textbf{2h} & \textbf{1h} & \textbf{30m} &
      \textbf{2h} & \textbf{1h} & \textbf{30m} &
      \textbf{2h} & \textbf{1h} & \textbf{30m} \\
    \hline
    Fault-free  &
       & 8.22 &  &
        & 8.05  &  &
        & 5.05  &  &
       & 7.07&  &
       & 5.89 &  &
       & 5.59 &  \\  
    Oobleck\cite{jang2023oobleck} &
       6.04 & 4.59 & -- &
       6.47 & 4.79 & -- &
       4.48 & 3.31 & -- &
       5.37 & 4.37 & -- &
       4.74 & 3.65 & -- &
       4.47 & 3.42 & -- \\    
    ReCycle\cite{gandhi2024recycle}&
       4.16 & 3.94 & -- &
       4.91 & 4.66 & -- &
       3.48 & 3.36 & -- &
       4.37 & 4.23 & -- &
       3.73 & 3.58 & -- &
       3.58 & 3.47 & -- \\
    \SysName{} &
       \textbf{7.55} & \textbf{6.48} & \textbf{5.46} &
       \textbf{7.95} & \textbf{7.23} & \textbf{5.95} &
       \textbf{4.80} & \textbf{4.38} & \textbf{3.53} &
       \textbf{6.57} & \textbf{5.95} & \textbf{5.62} &
       \textbf{5.59} & \textbf{5.06} & \textbf{4.45} &
       \textbf{4.80} & \textbf{4.23} & \textbf{3.86} \\
    \noalign{\hrule height 1pt}
  \end{tabular*}
  \label{tab:training-throughput-wide}
\end{table*}

\begin{figure*}
  \centering
  \includegraphics[width=1\linewidth]{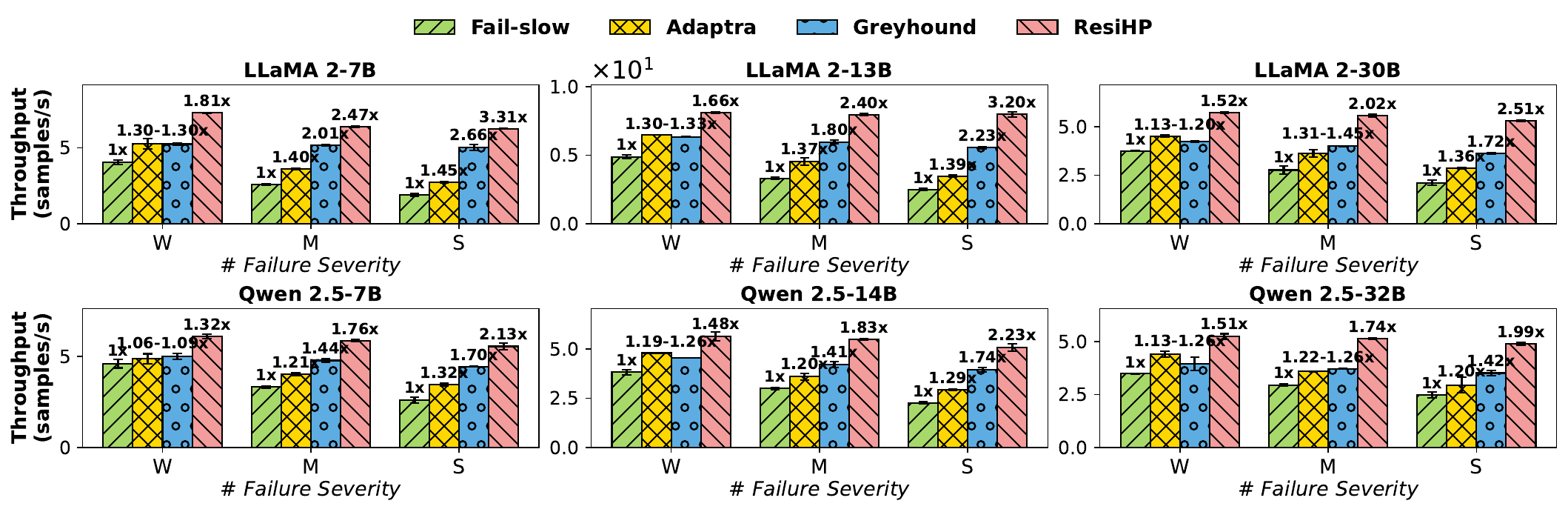}
    \vspace{-20pt}
  \caption{\gw{Effectiveness of the \texttt{Scheduler} in mitigating various fail-slow severities.}}
  \label{fig:various_fail-slow_severities}
  \vspace{-5pt}
\end{figure*}

\noindent\textbf{Time Prediction Accuracy.}
\gw{We first assess the micro-batch and iteration-time predictors across three representative training setups varying in sequence lengths, model sizes, and pipeline schedules. As shown in Table~\ref{tab:prediction-accuracy}, the micro-batch time predictor achieves a MAPE of 1.19\%--1.58\%, while the iteration-time predictor achieves 2.81\%--5.06\%. \texttt{Detector} accurately estimates healthy execution times even under substantial workload variability, providing a reliable baseline to filter out the benign iteration-time fluctuations.}

\noindent\textbf{Failure Detection Accuracy and Overhead.}
\gw{To evaluate fail-slow detection accuracy, we follow Greyhound~\cite{wu2025greyhound} and launch multiple short training jobs. In approximately half of the jobs, we inject a persistent fail-slow at a random iteration after the first 50 iterations and before the last 50 iterations, leaving sufficient time for detector warm-up and failure response. Table~\ref{tab:false-alarms} shows that the \texttt{Detector} substantially reduces false alarms and the associated overhead relative to Greyhound. Across all evaluated setups, the \texttt{Detector} incurs only 34--49 ms of false-alarm overhead, compared with 1.24--1.72 s for Greyhound. This reduction comes from filtering benign workload-induced spikes before invoking the validation phase. In addition, the \texttt{Detector} achieves over 99.0\% accuracy in identifying fail-slow anomalies and 99.6\% accuracy for fail-stop failures, demonstrating its effectiveness and robustness in failure diagnosis. This demonstrates that the \texttt{Detector} reliably covers fail-stop and persistent fail-slow failures, while failures without liveness or timing signatures are outside the current scope.}

\subsection{Scheduler Effectiveness}\label{sec:effective}
\noindent\textbf{Comparison against ReCycle\cite{gandhi2024recycle} and Oobleck\cite{jang2023oobleck} under Fail-stop Failures.} 
\gw{
To evaluate \SysName{}'s adaptability across a wide spectrum of deployment scenarios, we vary the fail-stop frequency from once every 2 hours to once every 30 minutes across 4--16h training sessions (scaled by model size), following prior work~\cite{gandhi2024recycle,jang2023oobleck,athlur2022varuna}. Workers are monotonically terminated over time, leaving only 50\% of the initial cluster intact in the 30m scenario.
}

\gw{Table~\ref{tab:training-throughput-wide} shows that ReCycle already suffers throughput drops of up to 49.4\% at the 2h frequency due to severe inter-DP imbalance induced by workload migration. Oobleck performs better initially, but still degrades by up to 44.2\% at the 1h frequency due to imbalanced heterogeneous pipelines and high reconfiguration latency.} Both baselines abort training under the severe 30m frequency, as they conservatively discard entire TP groups after intra-group failures and cannot recover when all DP replicas of a pipeline stage fail. In contrast, \SysName{} sustains training even in the 30m scenario. Across the comparable 2h and 1h cases, it achieves 1.22--1.82$\times$ and 1.07--1.51$\times$ throughput speedups over ReCycle and Oobleck, respectively, by progressively adapting TP, PP, and DP to salvage fragmented resources and maintain workload balance.

\noindent\textbf{Comparison against Greyhound~\cite{wu2025greyhound} and Adaptra~\cite{wu2025adaptra} under Fail-slow Failures.} 
\gw{To evaluate \SysName{}'s ability to handle realistic fail-slow scenarios of varying severities, we inject weak (W), medium (M), and severe (S) fail-slow failures. Following experimental setups in prior work~\cite{wu2025greyhound,wu2025adaptra}, these fail-slow categories induce unmitigated throughput drops of roughly 35\%, 55\%, and 70\% relative to the fault-free baseline, respectively. As illustrated in Figure~\ref{fig:various_fail-slow_severities}, \SysName{} improves the overall throughput of the degraded baseline by 1.32--3.31$\times$, translating to 1.18--2.30$\times$ and 1.22--1.46$\times$ speedups over Adaptra and Greyhound, respectively.} By consistently realigning workloads to match heterogeneous device speeds, \SysName{} maintains high training efficiency across diverse pipeline configurations and fail-slow severities.

\noindent\textbf{Handling mixed Failures.}
To evaluate effectiveness under complex and realistic failure scenarios, we alternately inject fail-stop failures and medium-severity fail-slow failures during training. \gw{Figure~\ref{fig:fail-slow&fail-stop} shows that \SysName{} improves throughput by 1.48--4.39$\times$, 1.22--4.32$\times$, and 1.04--3.57$\times$ over ReCycle, strengthened ReCycle, and strengthened Oobleck, respectively.} Notably, strengthened ReCycle provides negligible benefit over its vanilla counterpart in these mixed-failure scenarios. This is because devices already degraded by fail-slow failures may also receive additional workloads from crashed DP peers, turning them into severe stragglers that throttle end-to-end throughput.

\begin{figure*}
  \centering
  \includegraphics[width=1\linewidth]{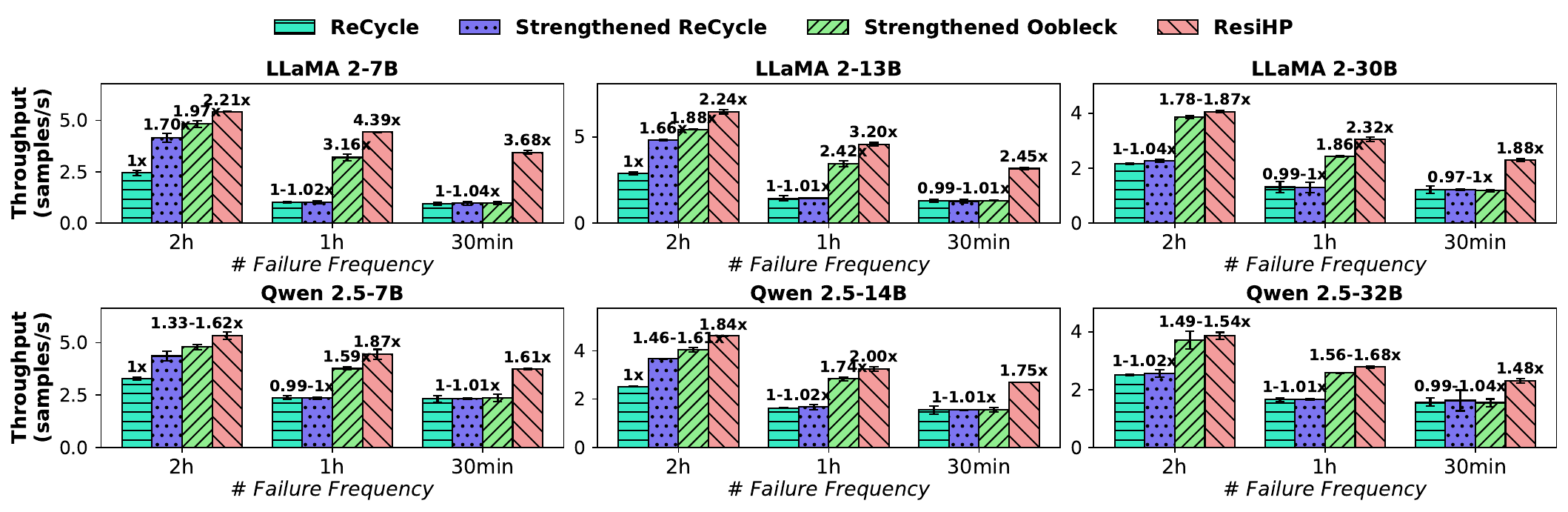}
    \vspace{-20pt}
  \caption{\gw{Effectiveness of the \texttt{Scheduler} in handling mixed failures.}}
  \label{fig:fail-slow&fail-stop}
\vspace{-5pt}
\end{figure*}

\begin{figure}[!t]
    \centering
    \includegraphics[width=1\linewidth]{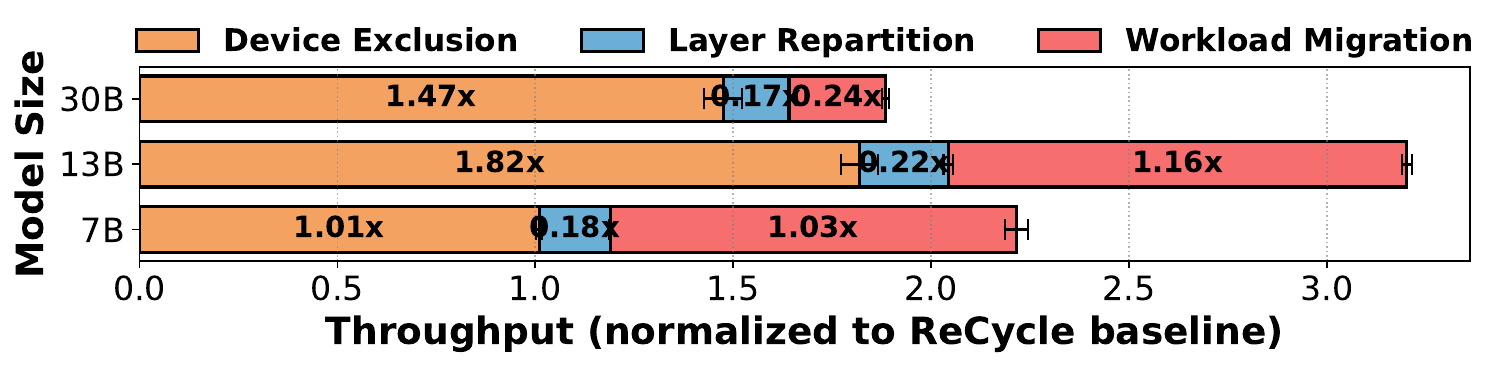}
        \vspace{-20pt}
    \caption{\gw{Performance breakdown for handling mixed failures across LLaMA 2-7B, 13B, and 30B models, with failure frequencies of 2h, 1h, and 30 min, respectively.}}
    \label{fig:braek_down}
    \vspace{-10pt}
\end{figure}

\noindent\textbf{Performance Ablation.}
\gw{We quantify the contribution of each \SysName{} component by incrementally enabling selective device exclusion, adaptive layer repartition, and workload migration. 
\autoref{fig:braek_down} reports each component's throughput contribution normalized to ReCycle. 
Among the three components, selective device exclusion provides the largest gain because it directly preserves useful TP computation capacity and reduces resource waste under failures. Adaptive layer repartition contributes less because its uniform application across DP replicas limits flexibility. In contrast, workload migration is more fine-grained, dynamically rebalancing workloads across DP replicas according to pipeline progress and residual performance skew. 
To further explain how \SysName{} mitigates the failure amplification in Figure~\ref{fig:amplification}, we isolate its impact at each propagation level. \SysName{} reduces the delay to 0.64$\times$ at the fail-slow node, 2.03$\times$ at TP, 8.72$\times$ at PP, and 11.14$\times$ at DP, showing that progressive adaptation limits failure propagation across dimensions and preserves high training efficiency.
}

\noindent\textbf{Validation of Fail-stop Handling on Convergence.} \gw{To ensure \SysName{} preserves training dynamics, we trained LLaMA 2-7B for 2,500 iterations under a fault-free baseline and \SysName{} with injected failures. Figure \ref{fig:training_loss} shows that both loss curves tightly overlap, sharing identical decay trends and final loss values. While a transient loss spike occurs during the third injected failure, the model immediately recovers because \SysName{} focuses on the system level without altering the mathematical semantics of training. Thus, \SysName{} can maintain strict convergence and final training quality without disrupting the overall loss trajectory.}

\begin{figure}[t!]
    \centering
    \includegraphics[width=1\linewidth]{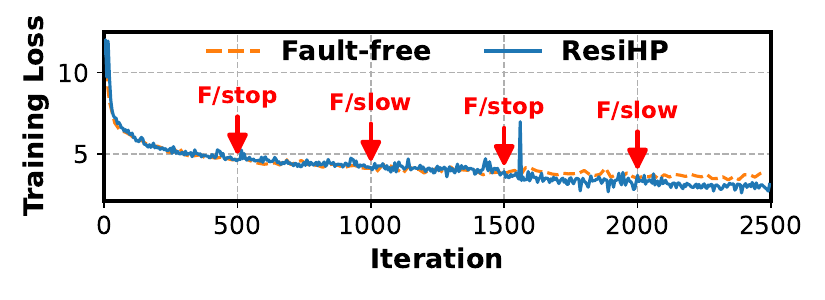}
    \vspace{-30pt}
    \caption{\gw{LLaMA 2-7B training loss: Fault-free baseline (orange) and recovered trajectory by ResiHP (blue).}}
    \label{fig:training_loss}
    \vspace{-10pt}
\end{figure}

\noindent\textbf{System Overhead.}
\gw{The system overhead consists of three main components: the \texttt{Detector} identifying failures, the \texttt{Scheduler} generating an adaptation plan, and the reconfiguration process modifying the 3D parallelism dimensions.
As Figure \ref{fig:overhead} demonstrates, the \texttt{Detector} overhead is negligible, increasing the per-iteration time by only 1.2--1.5\%. The warm-up profiling for the linear FLOPs-to-time model is a one-time cost which is not included here.
The \texttt{Scheduler}'s planning overhead scales with model size due to larger PP degrees but remains minimal, taking just 1.44s for the 32B model (less than half a training iteration).
Communication group reconstruction overhead is bounded to under 2s across all three models. 
Lastly, layer transfer overhead during reconfiguration scales with model size and transfer volume, but can be effectively amortized over long-running training.}

\begin{figure}[t]
    \centering
    \includegraphics[width=1\linewidth]{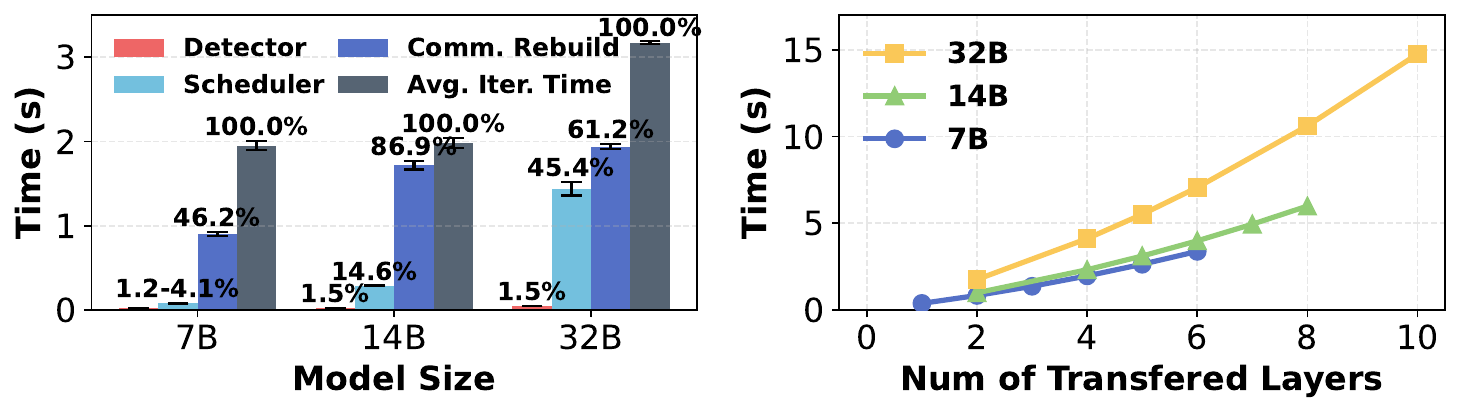}
    \vspace{-20pt}
 \caption{\gw{\textbf{Left}: Overhead of \SysName{} across Qwen 2.5 models. \textbf{Right}: Overhead of layer transfer during reconfiguration across Qwen 2.5 models.}}
  \label{fig:overhead}
  \vspace{-10pt}
\end{figure}

\subsection{Large-Scale Evaluation}
\label{sec:scale}

To evaluate \SysName{} at scale, we train LLaMA 2-70B on 256 NVIDIA A100 GPUs with $(TP, DP, PP)=(4,4,16)$. We use a dynamic training scenario with recurring failures and re-joins to evaluate end-to-end detection and mitigation.

\noindent\textbf{Detection.}
We first evaluate the \texttt{Detector}. As shown in Figure~\ref{fig:end-to-end}, throughput drops align closely with failure occurrences, and the subsequent rebounds indicate successful failure detection and mitigation. \texttt{Detector} identifies all failures within 2--3 training iterations despite high failure concurrency. When fail-stop and fail-slow failures co-occur, \texttt{Detector} prioritizes fail-stop detection to restore training first; after \texttt{Scheduler} reconfigures the job, \texttt{Detector} resumes detecting the remaining fail-slow failures.

\noindent\textbf{Failure Handling.}
Figure~\ref{fig:end-to-end} shows that strengthened ReCycle becomes increasingly ineffective as mixed failures accumulate over time, especially around iterations 2000 and 6000. This is because it may reassign workloads from failed DP peers to devices already degraded by fail-slow failures, further turning them into severe bottlenecks that significantly slow the entire cluster.

By combining accurate detection with effective mitigation, \SysName{} substantially reduces the impact of both fail-stop and fail-slow failures throughout training. \gw{In terms of average end-to-end throughput, \SysName{} achieves 1.39$\times$ and 1.11$\times$ speedups over strengthened ReCycle and strengthened Oobleck, respectively.}

\UCsection{Related Work}
\noindent\textbf{System Failure Analysis.}
Failure analysis has been widely studied in cloud services~\cite{chow2024servicelab,gan2021sage,gan2019seer}, operating systems~\cite{zhang2024illuminating}, and storage systems~\cite{gunawi2018fail,PERSEUS}. LLM training differs by using thousands of expensive GPUs under tightly synchronized execution, where a single failed or slow component can stall the entire job and recovery often requires coordinated cluster-wide reconfiguration.
\gw{Silent data corruptions (SDCs) represent another important failure class. Unlike fail-stop and fail-slow failures, SDCs may not affect device liveness or execution time, and thus require different signals. Prior works suggest lightweight checks on critical layers during LLM inference~\cite{sun2025ft2} and training-level anomalies such as loss spikes or parameter drift~\cite{ma2025understanding}. Extending \SysName{} with an SDC-specific detector based on these signals could reuse our system-level adaptations to isolate suspicious devices or stages, and we leave this to future work.}

\noindent\textbf{Data Heterogeneity.}
Variable-length sequences in LLM training induce workload variability across micro-batches and can further cause load imbalance under parallel execution~\cite{jiang2024dynapipe, FlexSP, ge2025bytescale, AttnIs}. Prior work mitigates such heterogeneity through dynamic scheduling~\cite{jiang2024dynapipe}, adaptive parallelism~\cite{FlexSP}, balanced data assignment~\cite{ge2025bytescale}, and data reordering~\cite{DistTrain}. However, these techniques do not explicitly stabilize the iteration-time series for failure detection, and residual fluctuations may still trigger spurious alarms.

\begin{figure}[t]
    \centering
    \includegraphics[width=1\linewidth]{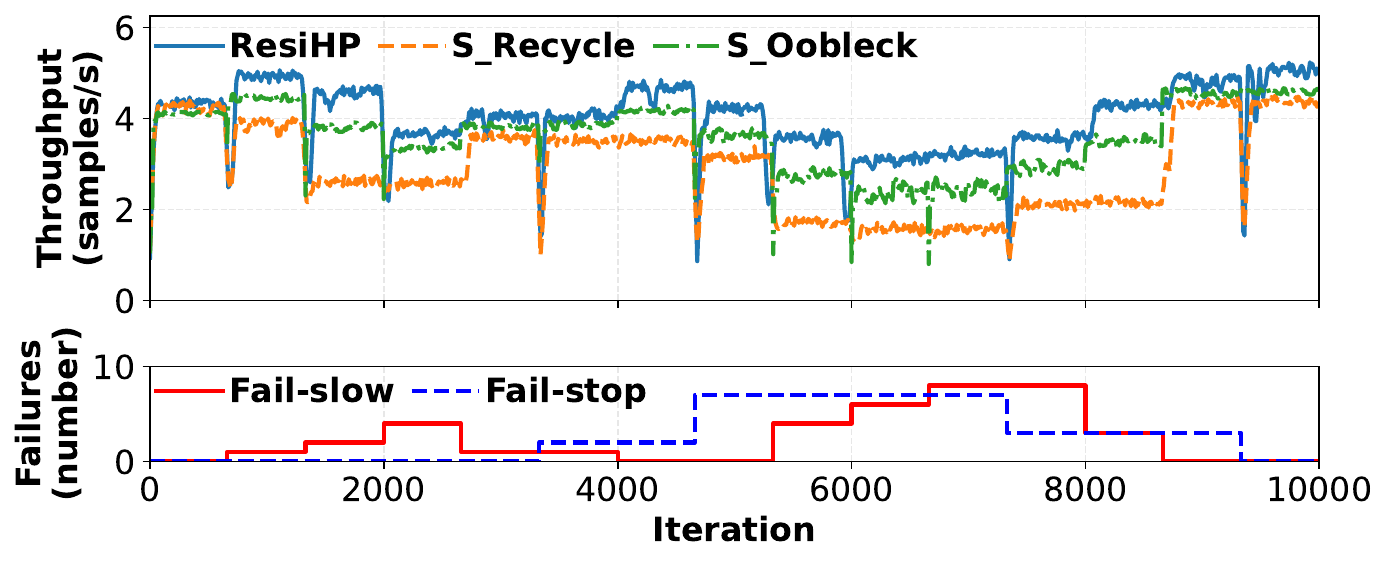}
    \vspace{-20pt}
    \caption{\gw{Evaluation of \SysName{} for a 256 A100 GPU training with fail-stop and fail-slow failures.}}
    \label{fig:end-to-end}
    \vspace{-10pt}
\end{figure}

\noindent\textbf{Resilient and Efficient LLM Training.} Prior systems improve resilience under failures, preemptions, and resource changes~\cite{Bamboo,jang2023oobleck,gandhi2024recycle,kang2025elaswave,ye2024survey}. These systems mainly target fail-stop failures or elasticity events, whereas \SysName{} also diagnoses and mitigates fail-slow performance degradation through progressive adaptations. Some systems automatically search for efficient parallel training plans~\cite{cai2022tensoropt,athlur2022varuna,zheng2022alpa,lin2024tessel,guo2025adaptis}. \SysName{} can benefit from these parallelism optimization strategies to further improve training efficiency. 
\UCsection{conclusion}
This paper presents \SysName{}, a system that automatically detects and mitigates both fail-slow and fail-stop failures in large-scale LLM training. \SysName{} combines a workload-aware execution time predictor that enables accurate fail-slow detection. Also, \SysName{} orchestrates a scheduler that jointly adapts hybrid parallelism via dynamic TP reconfiguration, layer repartition, and adaptive progress-aware workload migration. Our evaluation demonstrates the efficiency and scalability of \SysName{}.

\bibliographystyle{ACM-Reference-Format}
\bibliography{reference}

@inproceedings{ge2025bytescale,
  title={ByteScale: Communication-Efficient Scaling of LLM Training with a 2048K Context Length on 16384 GPUs},
  author={Ge, Hao and Feng, Junda and Huang, Qi and Fu, Fangcheng and Nie, Xiaonan and Zuo, Lei and Lin, Haibin and Cui, Bin and Liu, Xin},
  booktitle={Proceedings of the ACM SIGCOMM 2025 Conference},
  pages={963--978},
  year={2025}
}

@inproceedings{jiang2024dynapipe,
  title={DynaPipe: Optimizing multi-task training through dynamic pipelines},
  author={Jiang, Chenyu and Jia, Zhen and Zheng, Shuai and Wang, Yida and Wu, Chuan},
  booktitle={Proceedings of the Nineteenth European Conference on Computer Systems},
  pages={542--559},
  year={2024}
}

@article{lepikhin2020gshard,
  title={Gshard: Scaling giant models with conditional computation and automatic sharding},
  author={Lepikhin, Dmitry and Lee, HyoukJoong and Xu, Yuanzhong and Chen, Dehao and Firat, Orhan and Huang, Yanping and Krikun, Maxim and Shazeer, Noam and Chen, Zhifeng},
  journal={arXiv preprint arXiv:2006.16668},
  year={2020}
}

@article{megatronSP,
  title={Reducing activation recomputation in large transformer models},
  author={Korthikanti, Vijay Anand and Casper, Jared and Lym, Sangkug and McAfee, Lawrence and Andersch, Michael and Shoeybi, Mohammad and Catanzaro, Bryan},
  journal={Proceedings of Machine Learning and Systems},
  volume={5},
  pages={341--353},
  year={2023}
}

@misc{DeepSpeedU,
      title={DeepSpeed Ulysses: System Optimizations for Enabling Training of Extreme Long Sequence Transformer Models}, 
      author={Sam Ade Jacobs and Masahiro Tanaka and Chengming Zhang and Minjia Zhang and Shuaiwen Leon Song and Samyam Rajbhandari and Yuxiong He},
      year={2023},
      eprint={2309.14509},
      archivePrefix={arXiv},
      primaryClass={cs.LG},
      url={https://arxiv.org/abs/2309.14509}, 
}

@article{shoeybi2019megatron,
  title={Megatron-lm: Training multi-billion parameter language models using model parallelism},
  author={Shoeybi, Mohammad and Patwary, Mostofa and Puri, Raul and LeGresley, Patrick and Casper, Jared and Catanzaro, Bryan},
  journal={arXiv preprint arXiv:1909.08053},
  year={2019}
}

@inproceedings{wu2025greyhound,
  title={$\{$GREYHOUND$\}$: Hunting $\{$Fail-Slows$\}$ in $\{$Hybrid-Parallel$\}$ Training at Scale},
  author={Wu, Tianyuan and Wang, Wei and Yu, Yinghao and Yang, Siran and Wu, Wenchao and Duan, Qinkai and Yang, Guodong and Wang, Jiamang and Qu, Lin and Zhang, Liping},
  booktitle={2025 USENIX Annual Technical Conference (USENIX ATC 25)},
  pages={731--747},
  year={2025}
}

@inproceedings{jiang2024megascale,
  title={$\{$MegaScale$\}$: Scaling large language model training to more than 10,000 $\{$GPUs$\}$},
  author={Jiang, Ziheng and Lin, Haibin and Zhong, Yinmin and Huang, Qi and Chen, Yangrui and Zhang, Zhi and Peng, Yanghua and Li, Xiang and Xie, Cong and Nong, Shibiao and others},
  booktitle={21st USENIX Symposium on Networked Systems Design and Implementation (NSDI 24)},
  pages={745--760},
  year={2024}
}

@inproceedings{jang2023oobleck,
  title={Oobleck: Resilient distributed training of large models using pipeline templates},
  author={Jang, Insu and Yang, Zhenning and Zhang, Zhen and Jin, Xin and Chowdhury, Mosharaf},
  booktitle={Proceedings of the 29th Symposium on Operating Systems Principles},
  pages={382--395},
  year={2023}
}

@inproceedings{gandhi2024recycle,
  title={Recycle: Resilient training of large dnns using pipeline adaptation},
  author={Gandhi, Swapnil and Zhao, Mark and Skiadopoulos, Athinagoras and Kozyrakis, Christos},
  booktitle={Proceedings of the ACM SIGOPS 30th Symposium on Operating Systems Principles},
  pages={211--228},
  year={2024}
}

@article{lin2025understanding,
  title={Understanding Stragglers in Large Model Training Using What-if Analysis},
  author={Lin, Jinkun and Jiang, Ziheng and Song, Zuquan and Zhao, Sida and Yu, Menghan and Wang, Zhanghan and Wang, Chenyuan and Shi, Zuocheng and Shi, Xiang and Jia, Wei and others},
  journal={arXiv preprint arXiv:2505.05713},
  year={2025}
}

@article{guo2025adaptis,
  title={AdaPtis: Reducing Pipeline Bubbles with Adaptive Pipeline Parallelism on Heterogeneous Models},
  author={Guo, Jihu and Ma, Tenghui and Gao, Wei and Sun, Peng and Li, Jiaxing and Chen, Xun and Jin, Yuyang and Lin, Dahua},
  journal={arXiv preprint arXiv:2509.23722},
  year={2025}
}

@article{wu2025adaptra,
  title={Adaptra: Straggler-Resilient Hybrid-Parallel Training with Pipeline Adaptation},
  author={Wu, Tianyuan and Cao, Lunxi and Lu, Hanfeng and Jiang, Xiaoxiao and Yu, Yinghao and Yang, Siran and Yang, Guodong and Wang, Jiamang and Qu, Lin and Zhang, Liping and others},
  journal={arXiv preprint arXiv:2504.19232},
  year={2025}
}

@article{workshop2022bloom,
  title={Bloom: A 176b-parameter open-access multilingual language model},
  author={Workshop, BigScience and Scao, Teven Le and Fan, Angela and Akiki, Christopher and Pavlick, Ellie and Ili{\'c}, Suzana and Hesslow, Daniel and Castagn{\'e}, Roman and Luccioni, Alexandra Sasha and Yvon, Fran{\c{c}}ois and others},
  journal={arXiv preprint arXiv:2211.05100},
  year={2022}
}

@article{dubey2024llama,
  title={The llama 3 herd of models},
  author={Dubey, Abhimanyu and Jauhri, Abhinav and Pandey, Abhinav and Kadian, Abhishek and Al-Dahle, Ahmad and Letman, Aiesha and Mathur, Akhil and Schelten, Alan and Yang, Amy and Fan, Angela and others},
  journal={arXiv e-prints},
  pages={arXiv--2407},
  year={2024}
}

@misc{openai2024gpt4technicalreport,
      title={GPT-4 Technical Report}, 
      author={OpenAI and Josh Achiam and Steven Adler and Sandhini Agarwal and Lama Ahmad and Ilge Akkaya and Florencia Leoni Aleman and Diogo Almeida and Janko Altenschmidt and Sam Altman and Shyamal Anadkat and Red Avila and Igor Babuschkin and Suchir Balaji and Valerie Balcom and Paul Baltescu and Haiming Bao and Mohammad Bavarian and Jeff Belgum and Irwan Bello and Jake Berdine and Gabriel Bernadett-Shapiro and Christopher Berner and Lenny Bogdonoff and Oleg Boiko and Madelaine Boyd and Anna-Luisa Brakman and Greg Brockman and Tim Brooks and Miles Brundage and Kevin Button and Trevor Cai and Rosie Campbell and Andrew Cann and Brittany Carey and Chelsea Carlson and Rory Carmichael and Brooke Chan and Che Chang and Fotis Chantzis and Derek Chen and Sully Chen and Ruby Chen and Jason Chen and Mark Chen and Ben Chess and Chester Cho and Casey Chu and Hyung Won Chung and Dave Cummings and Jeremiah Currier and Yunxing Dai and Cory Decareaux and Thomas Degry and Noah Deutsch and Damien Deville and Arka Dhar and David Dohan and Steve Dowling and Sheila Dunning and Adrien Ecoffet and Atty Eleti and Tyna Eloundou and David Farhi and Liam Fedus and Niko Felix and Simón Posada Fishman and Juston Forte and Isabella Fulford and Leo Gao and Elie Georges and Christian Gibson and Vik Goel and Tarun Gogineni and Gabriel Goh and Rapha Gontijo-Lopes and Jonathan Gordon and Morgan Grafstein and Scott Gray and Ryan Greene and Joshua Gross and Shixiang Shane Gu and Yufei Guo and Chris Hallacy and Jesse Han and Jeff Harris and Yuchen He and Mike Heaton and Johannes Heidecke and Chris Hesse and Alan Hickey and Wade Hickey and Peter Hoeschele and Brandon Houghton and Kenny Hsu and Shengli Hu and Xin Hu and Joost Huizinga and Shantanu Jain and Shawn Jain and Joanne Jang and Angela Jiang and Roger Jiang and Haozhun Jin and Denny Jin and Shino Jomoto and Billie Jonn and Heewoo Jun and Tomer Kaftan and Łukasz Kaiser and Ali Kamali and Ingmar Kanitscheider and Nitish Shirish Keskar and Tabarak Khan and Logan Kilpatrick and Jong Wook Kim and Christina Kim and Yongjik Kim and Jan Hendrik Kirchner and Jamie Kiros and Matt Knight and Daniel Kokotajlo and Łukasz Kondraciuk and Andrew Kondrich and Aris Konstantinidis and Kyle Kosic and Gretchen Krueger and Vishal Kuo and Michael Lampe and Ikai Lan and Teddy Lee and Jan Leike and Jade Leung and Daniel Levy and Chak Ming Li and Rachel Lim and Molly Lin and Stephanie Lin and Mateusz Litwin and Theresa Lopez and Ryan Lowe and Patricia Lue and Anna Makanju and Kim Malfacini and Sam Manning and Todor Markov and Yaniv Markovski and Bianca Martin and Katie Mayer and Andrew Mayne and Bob McGrew and Scott Mayer McKinney and Christine McLeavey and Paul McMillan and Jake McNeil and David Medina and Aalok Mehta and Jacob Menick and Luke Metz and Andrey Mishchenko and Pamela Mishkin and Vinnie Monaco and Evan Morikawa and Daniel Mossing and Tong Mu and Mira Murati and Oleg Murk and David Mély and Ashvin Nair and Reiichiro Nakano and Rajeev Nayak and Arvind Neelakantan and Richard Ngo and Hyeonwoo Noh and Long Ouyang and Cullen O'Keefe and Jakub Pachocki and Alex Paino and Joe Palermo and Ashley Pantuliano and Giambattista Parascandolo and Joel Parish and Emy Parparita and Alex Passos and Mikhail Pavlov and Andrew Peng and Adam Perelman and Filipe de Avila Belbute Peres and Michael Petrov and Henrique Ponde de Oliveira Pinto and Michael and Pokorny and Michelle Pokrass and Vitchyr H. Pong and Tolly Powell and Alethea Power and Boris Power and Elizabeth Proehl and Raul Puri and Alec Radford and Jack Rae and Aditya Ramesh and Cameron Raymond and Francis Real and Kendra Rimbach and Carl Ross and Bob Rotsted and Henri Roussez and Nick Ryder and Mario Saltarelli and Ted Sanders and Shibani Santurkar and Girish Sastry and Heather Schmidt and David Schnurr and John Schulman and Daniel Selsam and Kyla Sheppard and Toki Sherbakov and Jessica Shieh and Sarah Shoker and Pranav Shyam and Szymon Sidor and Eric Sigler and Maddie Simens and Jordan Sitkin and Katarina Slama and Ian Sohl and Benjamin Sokolowsky and Yang Song and Natalie Staudacher and Felipe Petroski Such and Natalie Summers and Ilya Sutskever and Jie Tang and Nikolas Tezak and Madeleine B. Thompson and Phil Tillet and Amin Tootoonchian and Elizabeth Tseng and Preston Tuggle and Nick Turley and Jerry Tworek and Juan Felipe Cerón Uribe and Andrea Vallone and Arun Vijayvergiya and Chelsea Voss and Carroll Wainwright and Justin Jay Wang and Alvin Wang and Ben Wang and Jonathan Ward and Jason Wei and CJ Weinmann and Akila Welihinda and Peter Welinder and Jiayi Weng and Lilian Weng and Matt Wiethoff and Dave Willner and Clemens Winter and Samuel Wolrich and Hannah Wong and Lauren Workman and Sherwin Wu and Jeff Wu and Michael Wu and Kai Xiao and Tao Xu and Sarah Yoo and Kevin Yu and Qiming Yuan and Wojciech Zaremba and Rowan Zellers and Chong Zhang and Marvin Zhang and Shengjia Zhao and Tianhao Zheng and Juntang Zhuang and William Zhuk and Barret Zoph},
      year={2024},
      eprint={2303.08774},
      archivePrefix={arXiv},
      primaryClass={cs.CL},
      url={https://arxiv.org/abs/2303.08774}, 
}

@misc{yang2025qwen3technicalreport,
      title={Qwen3 Technical Report}, 
      author={An Yang and Anfeng Li and Baosong Yang and Beichen Zhang and Binyuan Hui and Bo Zheng and Bowen Yu and Chang Gao and Chengen Huang and Chenxu Lv and Chujie Zheng and Dayiheng Liu and Fan Zhou and Fei Huang and Feng Hu and Hao Ge and Haoran Wei and Huan Lin and Jialong Tang and Jian Yang and Jianhong Tu and Jianwei Zhang and Jianxin Yang and Jiaxi Yang and Jing Zhou and Jingren Zhou and Junyang Lin and Kai Dang and Keqin Bao and Kexin Yang and Le Yu and Lianghao Deng and Mei Li and Mingfeng Xue and Mingze Li and Pei Zhang and Peng Wang and Qin Zhu and Rui Men and Ruize Gao and Shixuan Liu and Shuang Luo and Tianhao Li and Tianyi Tang and Wenbiao Yin and Xingzhang Ren and Xinyu Wang and Xinyu Zhang and Xuancheng Ren and Yang Fan and Yang Su and Yichang Zhang and Yinger Zhang and Yu Wan and Yuqiong Liu and Zekun Wang and Zeyu Cui and Zhenru Zhang and Zhipeng Zhou and Zihan Qiu},
      year={2025},
      eprint={2505.09388},
      archivePrefix={arXiv},
      primaryClass={cs.CL},
      url={https://arxiv.org/abs/2505.09388}, 
}

@article{FailSlow,
author = {Gunawi, Haryadi S. and Suminto, Riza O. and Sears, Russell and Golliher, Casey and Sundararaman, Swaminathan and Lin, Xing and Emami, Tim and Sheng, Weiguang and Bidokhti, Nematollah and McCaffrey, Caitie and Srinivasan, Deepthi and Panda, Biswaranjan and Baptist, Andrew and Grider, Gary and Fields, Parks M. and Harms, Kevin and Ross, Robert B. and Jacobson, Andree and Ricci, Robert and Webb, Kirk and Alvaro, Peter and Runesha, H. Birali and Hao, Mingzhe and Li, Huaicheng},
title = {Fail-Slow at Scale: Evidence of Hardware Performance Faults in Large Production Systems},
year = {2018},
issue_date = {August 2018},
publisher = {Association for Computing Machinery},
address = {New York, NY, USA},
volume = {14},
number = {3},
issn = {1553-3077},
url = {https://doi.org/10.1145/3242086},
doi = {10.1145/3242086},
journal = {ACM Trans. Storage},
month = oct,
articleno = {23},
numpages = {26}
}

@inproceedings{SuperBench,
author = {Xiong, Yifan and Jiang, Yuting and Yang, Ziyue and Qu, Lei and Zhao, Guoshuai and Liu, Shuguang and Zhong, Dong and Pinzur, Boris and Zhang, Jie and Wang, Yang and Jose, Jithin and Pourreza, Hossein and Baxter, Jeff and Datta, Kushal and Ram, Prabhat and Melton, Luke and Chau, Joe and Cheng, Peng and Xiong, Yongqiang and Zhou, Lidong},
title = {SuperBench: improving cloud AI infrastructure reliability with proactive validation},
year = {2024},
isbn = {978-1-939133-41-0},
publisher = {USENIX Association},
address = {USA},
booktitle = {Proceedings of the 2024 USENIX Conference on Usenix Annual Technical Conference},
articleno = {51},
numpages = {16},
location = {Santa Clara, CA, USA},
series = {USENIX ATC'24}
}

@misc{OPT,
      title={OPT: Open Pre-trained Transformer Language Models}, 
      author={Susan Zhang and Stephen Roller and Naman Goyal and Mikel Artetxe and Moya Chen and Shuohui Chen and Christopher Dewan and Mona Diab and Xian Li and Xi Victoria Lin and Todor Mihaylov and Myle Ott and Sam Shleifer and Kurt Shuster and Daniel Simig and Punit Singh Koura and Anjali Sridhar and Tianlu Wang and Luke Zettlemoyer},
      year={2022},
      eprint={2205.01068},
      archivePrefix={arXiv},
      primaryClass={cs.CL},
      url={https://arxiv.org/abs/2205.01068}, 
}

@inproceedings{DistTrain,
author = {Zhang, Zili and Zhong, Yinmin and Jiang, Yimin and Hu, Hanpeng and Sun, Jianjian and Ge, Zheng and Zhu, Yibo and Jiang, Daxin and Jin, Xin},
title = {DistTrain: Addressing Model and Data Heterogeneity with Disaggregated Training for Multimodal Large Language Models},
year = {2025},
isbn = {9798400715242},
publisher = {Association for Computing Machinery},
address = {New York, NY, USA},
url = {https://doi.org/10.1145/3718958.3750472},
doi = {10.1145/3718958.3750472},
booktitle = {Proceedings of the ACM SIGCOMM 2025 Conference},
pages = {24–38},
numpages = {15},
location = {S\~{a}o Francisco Convent, Coimbra, Portugal},
series = {SIGCOMM '25}
}

@misc{FlexSP,
      title={FlexSP: Accelerating Large Language Model Training via Flexible Sequence Parallelism}, 
      author={Yujie Wang and Shiju Wang and Shenhan Zhu and Fangcheng Fu and Xinyi Liu and Xuefeng Xiao and Huixia Li and Jiashi Li and Faming Wu and Bin Cui},
      year={2025},
      eprint={2412.01523},
      archivePrefix={arXiv},
      primaryClass={cs.DC},
      url={https://arxiv.org/abs/2412.01523}, 
}

@article{Hydraulis,
author = {Li, Haoyang and Fu, Fangcheng and Lin, Sheng and Ge, Hao and Wang, Xuanyu and Niu, Jiawen and Xue, Jinbao and Tao, Yangyu and Wang, Di and Jiang, Jie and Cui, Bin},
title = {Hydraulis: Balancing Large Transformer Model Training via Co-designing Parallel Strategies and Data Assignment},
year = {2025},
issue_date = {December 2025},
publisher = {Association for Computing Machinery},
address = {New York, NY, USA},
volume = {3},
number = {6},
url = {https://doi.org/10.1145/3769802},
doi = {10.1145/3769802},
journal = {Proc. ACM Manag. Data},
month = dec,
articleno = {337},
numpages = {30}
}

@misc{AttnIs,
      title={Attention Is All You Need}, 
      author={Ashish Vaswani and Noam Shazeer and Niki Parmar and Jakob Uszkoreit and Llion Jones and Aidan N. Gomez and Lukasz Kaiser and Illia Polosukhin},
      year={2023},
      eprint={1706.03762},
      archivePrefix={arXiv},
      primaryClass={cs.CL},
      url={https://arxiv.org/abs/1706.03762}, 
}

@inproceedings{megatron2,
author = {Narayanan, Deepak and Shoeybi, Mohammad and Casper, Jared and LeGresley, Patrick and Patwary, Mostofa and Korthikanti, Vijay and Vainbrand, Dmitri and Kashinkunti, Prethvi and Bernauer, Julie and Catanzaro, Bryan and Phanishayee, Amar and Zaharia, Matei},
title = {Efficient large-scale language model training on GPU clusters using megatron-LM},
year = {2021},
isbn = {9781450384421},
publisher = {Association for Computing Machinery},
address = {New York, NY, USA},
url = {https://doi.org/10.1145/3458817.3476209},
doi = {10.1145/3458817.3476209},
booktitle = {Proceedings of the International Conference for High Performance Computing, Networking, Storage and Analysis},
articleno = {58},
numpages = {15},
location = {St. Louis, Missouri},
series = {SC '21}
}

@misc{SeqPacking,
      title={Efficient Sequence Packing without Cross-contamination: Accelerating Large Language Models without Impacting Performance}, 
      author={Mario Michael Krell and Matej Kosec and Sergio P. Perez and Andrew Fitzgibbon},
      year={2022},
      eprint={2107.02027},
      archivePrefix={arXiv},
      primaryClass={cs.CL},
      url={https://arxiv.org/abs/2107.02027}, 
}

@inproceedings{WLB-LLM,
author = {Wang, Zheng and Cai, Anna and Xie, Xinfeng and Pan, Zaifeng and Guan, Yue and Chu, Weiwei and Wang, Jie and Li, Shikai and Huang, Jianyu and Cai, Chris and Hao, Yuchen and Ding, Yufei},
title = {WLB-LLM: workload-balanced 4D parallelism for large language model training},
year = {2025},
isbn = {978-1-939133-47-2},
publisher = {USENIX Association},
address = {USA},
booktitle = {Proceedings of the 19th USENIX Conference on Operating Systems Design and Implementation},
articleno = {43},
numpages = {17},
location = {Boston, MA, USA},
series = {OSDI '25}
}

@inproceedings{PERSEUS,
author = {Lu, Ruiming and Xu, Erci and Zhang, Yiming and Zhu, Fengyi and Zhu, Zhaosheng and Wang, Mengtian and Zhu, Zongpeng and Xue, Guangtao and Shu, Jiwu and Li, Minglu and Wu, Jiesheng},
title = {PERSEUS: a fail-slow detection framework for cloud storage systems},
year = {2023},
isbn = {978-1-939133-32-8},
publisher = {USENIX Association},
address = {USA},
booktitle = {Proceedings of the 21st USENIX Conference on File and Storage Technologies},
articleno = {4},
numpages = {15},
location = {Santa Clara, CA, USA},
series = {FAST'23}
}

@inproceedings{Understandfailslow,
author = {Dong, Gen and Hua, Yu and Zhang, Yongle and Chen, Zhangyu and Chen, Menglei},
title = {Understanding and detecting fail-slow hardware failure bugs in cloud systems},
year = {2025},
isbn = {978-1-939133-48-9},
publisher = {USENIX Association},
address = {USA},
booktitle = {Proceedings of the 2025 USENIX Conference on Usenix Annual Technical Conference},
articleno = {66},
numpages = {16},
location = {Boston, MA, USA},
series = {USENIX ATC '25}
}

@misc{GitHub,
  author       = {{Hugging Face Datasets}},
  title        = {GitHub Code Dataset (codeparrot/github-code)},
  year         = {2023},
  howpublished = {\url{https://huggingface.co/datasets/codeparrot/github-code}},
  note         = {Accessed: 2025-12-XX},
}

@inproceedings {Bamboo,
author = {John Thorpe and Pengzhan Zhao and Jonathan Eyolfson and Yifan Qiao and Zhihao Jia and Minjia Zhang and Ravi Netravali and Guoqing Harry Xu},
title = {Bamboo: Making Preemptible Instances Resilient for Affordable Training of Large {DNNs}},
booktitle = {20th USENIX Symposium on Networked Systems Design and Implementation (NSDI 23)},
year = {2023},
isbn = {978-1-939133-33-5},
address = {Boston, MA},
pages = {497--513},
url = {https://www.usenix.org/conference/nsdi23/presentation/thorpe},
publisher = {USENIX Association},
month = apr
}

@misc{dsv3,
      title={DeepSeek-V3 Technical Report}, 
      author={DeepSeek-AI and Aixin Liu and Bei Feng and Bing Xue and Bingxuan Wang and Bochao Wu and Chengda Lu and Chenggang Zhao and Chengqi Deng and Chenyu Zhang and Chong Ruan and Damai Dai and Daya Guo and Dejian Yang and Deli Chen and Dongjie Ji and Erhang Li and Fangyun Lin and Fucong Dai and Fuli Luo and Guangbo Hao and Guanting Chen and Guowei Li and H. Zhang and Han Bao and Hanwei Xu and Haocheng Wang and Haowei Zhang and Honghui Ding and Huajian Xin and Huazuo Gao and Hui Li and Hui Qu and J. L. Cai and Jian Liang and Jianzhong Guo and Jiaqi Ni and Jiashi Li and Jiawei Wang and Jin Chen and Jingchang Chen and Jingyang Yuan and Junjie Qiu and Junlong Li and Junxiao Song and Kai Dong and Kai Hu and Kaige Gao and Kang Guan and Kexin Huang and Kuai Yu and Lean Wang and Lecong Zhang and Lei Xu and Leyi Xia and Liang Zhao and Litong Wang and Liyue Zhang and Meng Li and Miaojun Wang and Mingchuan Zhang and Minghua Zhang and Minghui Tang and Mingming Li and Ning Tian and Panpan Huang and Peiyi Wang and Peng Zhang and Qiancheng Wang and Qihao Zhu and Qinyu Chen and Qiushi Du and R. J. Chen and R. L. Jin and Ruiqi Ge and Ruisong Zhang and Ruizhe Pan and Runji Wang and Runxin Xu and Ruoyu Zhang and Ruyi Chen and S. S. Li and Shanghao Lu and Shangyan Zhou and Shanhuang Chen and Shaoqing Wu and Shengfeng Ye and Shengfeng Ye and Shirong Ma and Shiyu Wang and Shuang Zhou and Shuiping Yu and Shunfeng Zhou and Shuting Pan and T. Wang and Tao Yun and Tian Pei and Tianyu Sun and W. L. Xiao and Wangding Zeng and Wanjia Zhao and Wei An and Wen Liu and Wenfeng Liang and Wenjun Gao and Wenqin Yu and Wentao Zhang and X. Q. Li and Xiangyue Jin and Xianzu Wang and Xiao Bi and Xiaodong Liu and Xiaohan Wang and Xiaojin Shen and Xiaokang Chen and Xiaokang Zhang and Xiaosha Chen and Xiaotao Nie and Xiaowen Sun and Xiaoxiang Wang and Xin Cheng and Xin Liu and Xin Xie and Xingchao Liu and Xingkai Yu and Xinnan Song and Xinxia Shan and Xinyi Zhou and Xinyu Yang and Xinyuan Li and Xuecheng Su and Xuheng Lin and Y. K. Li and Y. Q. Wang and Y. X. Wei and Y. X. Zhu and Yang Zhang and Yanhong Xu and Yanhong Xu and Yanping Huang and Yao Li and Yao Zhao and Yaofeng Sun and Yaohui Li and Yaohui Wang and Yi Yu and Yi Zheng and Yichao Zhang and Yifan Shi and Yiliang Xiong and Ying He and Ying Tang and Yishi Piao and Yisong Wang and Yixuan Tan and Yiyang Ma and Yiyuan Liu and Yongqiang Guo and Yu Wu and Yuan Ou and Yuchen Zhu and Yuduan Wang and Yue Gong and Yuheng Zou and Yujia He and Yukun Zha and Yunfan Xiong and Yunxian Ma and Yuting Yan and Yuxiang Luo and Yuxiang You and Yuxuan Liu and Yuyang Zhou and Z. F. Wu and Z. Z. Ren and Zehui Ren and Zhangli Sha and Zhe Fu and Zhean Xu and Zhen Huang and Zhen Zhang and Zhenda Xie and Zhengyan Zhang and Zhewen Hao and Zhibin Gou and Zhicheng Ma and Zhigang Yan and Zhihong Shao and Zhipeng Xu and Zhiyu Wu and Zhongyu Zhang and Zhuoshu Li and Zihui Gu and Zijia Zhu and Zijun Liu and Zilin Li and Ziwei Xie and Ziyang Song and Ziyi Gao and Zizheng Pan},
      year={2025},
      eprint={2412.19437},
      archivePrefix={arXiv},
      primaryClass={cs.CL},
      url={https://arxiv.org/abs/2412.19437}, 
}

@inproceedings {IASO,
author = {Biswaranjan Panda and Deepthi Srinivasan and Huan Ke and Karan Gupta and Vinayak Khot and Haryadi S. Gunawi},
title = {{IASO}: A {Fail-Slow} Detection and Mitigation Framework for Distributed Storage Services},
booktitle = {2019 USENIX Annual Technical Conference (USENIX ATC 19)},
year = {2019},
isbn = {978-1-939133-03-8},
address = {Renton, WA},
pages = {47--62},
url = {https://www.usenix.org/conference/atc19/presentation/panda},
publisher = {USENIX Association},
month = jul
}

@inbook{GPipe,
author = {Huang, Yanping and Cheng, Youlong and Bapna, Ankur and Firat, Orhan and Chen, Mia Xu and Chen, Dehao and Lee, HyoukJoong and Ngiam, Jiquan and Le, Quoc V. and Wu, Yonghui and Chen, Zhifeng},
title = {GPipe: efficient training of giant neural networks using pipeline parallelism},
year = {2019},
publisher = {Curran Associates Inc.},
address = {Red Hook, NY, USA},
booktitle = {Proceedings of the 33rd International Conference on Neural Information Processing Systems},
articleno = {10},
numpages = {10}
}

@article{DP,
author = {Valiant, Leslie G.},
title = {A bridging model for parallel computation},
year = {1990},
issue_date = {Aug. 1990},
publisher = {Association for Computing Machinery},
address = {New York, NY, USA},
volume = {33},
number = {8},
issn = {0001-0782},
url = {https://doi.org/10.1145/79173.79181},
doi = {10.1145/79173.79181},
journal = {Commun. ACM},
month = aug,
pages = {103–111},
numpages = {9}
}

@inproceedings{Characterization,
author = {Hu, Qinghao and Ye, Zhisheng and Wang, Zerui and Wang, Guoteng and Zhang, Meng and Chen, Qiaoling and Sun, Peng and Lin, Dahua and Wang, Xiaolin and Luo, Yingwei and Wen, Yonggang and Zhang, Tianwei},
title = {Characterization of large language model development in the datacenter},
year = {2024},
isbn = {978-1-939133-39-7},
publisher = {USENIX Association},
address = {USA},
booktitle = {Proceedings of the 21st USENIX Symposium on Networked Systems Design and Implementation},
articleno = {39},
numpages = {21},
location = {Santa Clara, CA, USA},
series = {NSDI'24}
}

@inproceedings{chow2024servicelab,
  title={$\{$ServiceLab$\}$: Preventing tiny performance regressions at hyperscale through $\{$Pre-Production$\}$ testing},
  author={Chow, Mike and Wang, Yang and Wang, William and Hailu, Ayichew and Bopardikar, Rohan and Zhang, Bin and Qu, Jialiang and Meisner, David and Sonawane, Santosh and Zhang, Yunqi and others},
  booktitle={18th USENIX Symposium on Operating Systems Design and Implementation (OSDI 24)},
  pages={545--562},
  year={2024}
}

@article{gan2021sage,
  title={Sage: Leveraging ml to diagnose unpredictable performance in cloud microservices},
  author={Gan, Yu and Liang, Mingyu and Dev, Sundar and Lo, David and Delimitrou, Christina},
  journal={arXiv preprint arXiv:2112.06263},
  year={2021}
}

@inproceedings{gan2019seer,
  title={Seer: Leveraging big data to navigate the complexity of performance debugging in cloud microservices},
  author={Gan, Yu and Zhang, Yanqi and Hu, Kelvin and Cheng, Dailun and He, Yuan and Pancholi, Meghna and Delimitrou, Christina},
  booktitle={Proceedings of the twenty-fourth international conference on architectural support for programming languages and operating systems},
  pages={19--33},
  year={2019}
}

@article{gunawi2018fail,
  title={Fail-slow at scale: Evidence of hardware performance faults in large production systems},
  author={Gunawi, Haryadi S and Suminto, Riza O and Sears, Russell and Golliher, Casey and Sundararaman, Swaminathan and Lin, Xing and Emami, Tim and Sheng, Weiguang and Bidokhti, Nematollah and McCaffrey, Caitie and others},
  journal={ACM Transactions on Storage (TOS)},
  volume={14},
  number={3},
  pages={1--26},
  year={2018},
  publisher={ACM New York, NY, USA}
}

@inproceedings{zhang2024illuminating,
  title={Illuminating the gray zone: Non-intrusive gray failure localization in server operating systems},
  author={Zhang, Shenglin and Zhao, Yongxin and Xiong, Xiao and Sun, Yongqian and Nie, Xiaohui and Zhang, Jiacheng and Wang, Fenglai and Zheng, Xian and Zhang, Yuzhi and Pei, Dan},
  booktitle={Companion Proceedings of the 32nd ACM International Conference on the Foundations of Software Engineering},
  pages={126--137},
  year={2024}
}

@inproceedings{athlur2022varuna,
  title={Varuna: scalable, low-cost training of massive deep learning models},
  author={Athlur, Sanjith and Saran, Nitika and Sivathanu, Muthian and Ramjee, Ramachandran and Kwatra, Nipun},
  booktitle={Proceedings of the Seventeenth European Conference on Computer Systems},
  pages={472--487},
  year={2022}
}

@article{gao2025rollpacker,
  title={Rollpacker: Mitigating long-tail rollouts for fast, synchronous rl post-training},
  author={Gao, Wei and Zhao, Yuheng and An, Dakai and Wu, Tianyuan and Cao, Lunxi and Xiong, Shaopan and Huang, Ju and Wang, Weixun and Yang, Siran and Su, Wenbo and others},
  journal={arXiv preprint arXiv:2509.21009},
  year={2025}
}

@inproceedings{sun2025ft2,
  title={Ft2: First-token-inspired online fault tolerance on critical layers for generative large language models},
  author={Sun, Yu and Zhu, Zhu and Mulpuru, Cherish and Gioiosa, Roberto and Zhang, Zhao and Fang, Bo and Yang, Lishan},
  booktitle={Proceedings of the 34th International Symposium on High-Performance Parallel and Distributed Computing},
  pages={1--14},
  year={2025}
}

@inproceedings{ma2025understanding,
  title={Understanding silent data corruption in LLM training},
  author={Ma, Jeffrey Jian and Pei, Hengzhi and Lausen, Leonard and Karypis, George},
  booktitle={Proceedings of the 63rd Annual Meeting of the Association for Computational Linguistics (Volume 1: Long Papers)},
  pages={20372--20394},
  year={2025}
}

@misc{yang2024qwen2technicalreport,
      title={Qwen2 Technical Report}, 
      author={An Yang and Baosong Yang and Binyuan Hui and Bo Zheng and Bowen Yu and Chang Zhou and Chengpeng Li and Chengyuan Li and Dayiheng Liu and Fei Huang and Guanting Dong and Haoran Wei and Huan Lin and Jialong Tang and Jialin Wang and Jian Yang and Jianhong Tu and Jianwei Zhang and Jianxin Ma and Jianxin Yang and Jin Xu and Jingren Zhou and Jinze Bai and Jinzheng He and Junyang Lin and Kai Dang and Keming Lu and Keqin Chen and Kexin Yang and Mei Li and Mingfeng Xue and Na Ni and Pei Zhang and Peng Wang and Ru Peng and Rui Men and Ruize Gao and Runji Lin and Shijie Wang and Shuai Bai and Sinan Tan and Tianhang Zhu and Tianhao Li and Tianyu Liu and Wenbin Ge and Xiaodong Deng and Xiaohuan Zhou and Xingzhang Ren and Xinyu Zhang and Xipin Wei and Xuancheng Ren and Xuejing Liu and Yang Fan and Yang Yao and Yichang Zhang and Yu Wan and Yunfei Chu and Yuqiong Liu and Zeyu Cui and Zhenru Zhang and Zhifang Guo and Zhihao Fan},
      year={2024},
      eprint={2407.10671},
      archivePrefix={arXiv},
      primaryClass={cs.CL},
      url={https://arxiv.org/abs/2407.10671}, 
}

@book{jain1991art,
  title={The Art of Computer Systems Performance Analysis: Techniques for Experimental Design, Measurement, Simulation, and Modeling},
  author={Jain, Raj},
  year={1991},
  publisher={Wiley}
}

@inproceedings{georges2007statistically,
  title={Statistically rigorous Java performance evaluation},
  author={Georges, Andy and Buytaert, Dries and Eeckhout, Lieven},
  booktitle={Proceedings of the 22nd Annual ACM SIGPLAN Conference on Object-Oriented Programming Systems and Applications},
  year={2007}
}

@inproceedings{narayanan2019pipedream,
  title={PipeDream: Generalized pipeline parallelism for DNN training},
  author={Narayanan, Deepak and Harlap, Aaron and Phanishayee, Amar and Seshadri, Vivek and Devanur, Nikhil R and Ganger, Gregory R and Gibbons, Phillip B and Zaharia, Matei},
  booktitle={Proceedings of the 27th ACM symposium on operating systems principles},
  pages={1--15},
  year={2019}
}

@inproceedings{qi2024zero,
  title={Zero bubble (almost) pipeline parallelism},
  author={Qi, Penghui and Wan, Xinyi and Huang, Guangxing and Lin, Min},
  booktitle={The Twelfth International Conference on Learning Representations},
  year={2024}
}

@article{cai2022tensoropt,
  title={TensorOpt: Exploring the Tradeoffs in Distributed DNN Training With Auto-Parallelism},
  author={Cai, Zhenkun and Yan, Xiao and Ma, Kaihao and Wu, Yidi and Huang, Yuzhen and Cheng, James and Su, Teng and Yu, Fan},
  journal={IEEE Transactions on Parallel and Distributed Systems},
  volume={33},
  number={8},
  pages={1967--1981},
  year={2022}
}

@inproceedings{zheng2022alpa,
  title={Alpa: Automating Inter- and Intra-Operator Parallelism for Distributed Deep Learning},
  author={Zheng, Lianmin and Li, Zhuohan and Zhang, Hao and Zhuang, Yonghao and Chen, Zhifeng and Huang, Yanping and Wang, Yida and Xu, Yuanzhong and Zhuo, Danyang and Gonzalez, Joseph E. and Stoica, Ion},
  booktitle={OSDI},
  year={2022}
}

@inproceedings{lin2024tessel,
  title={Tessel: Boosting Distributed Execution of Large DNN Models via Flexible Schedule Search},
  author={Lin, Zhiqi and Miao, Youshan and Xu, Guanbin and Li, Cheng and Saarikivi, Olli and Maleki, Saeed and Yang, Fan},
  booktitle={HPCA},
  year={2024}
}

@article{kang2025elaswave,
  title={ElasWave: An Elastic-Native System for Scalable Hybrid-Parallel Training},
  author={Kang, Xueze and Xiang, Guangyu and Wang, Yuxin and Zhang, Hao and Fang, Yuchu and Zhou, Yuhang and Tang, Zhenheng and Lv, Youhui and Maman, Eliran and Wasserman, Mark and others},
  journal={arXiv preprint arXiv:2510.00606},
  year={2025}
}

@inproceedings{agudelo2020bayesian,
  title={Bayesian online prediction of change points},
  author={Agudelo-Espa{\~n}a, Diego and Gomez-Gonzalez, Sebastian and Bauer, Stefan and Sch{\"o}lkopf, Bernhard and Peters, Jan},
  booktitle={Conference on uncertainty in artificial intelligence},
  pages={320--329},
  year={2020},
  organization={PMLR}
}

@article{ye2024survey,
author = {Ye, Zhisheng and Gao, Wei and Hu, Qinghao and Sun, Peng and Wang, Xiaolin and Luo, Yingwei and Zhang, Tianwei and Wen, Yonggang},
title = {Deep Learning Workload Scheduling in GPU Datacenters: A Survey},
year = {2024},
volume = {56},
number = {6},
journal = {ACM Comput. Surv.},
month = jan,
articleno = {146},
numpages = {38}
}

@article{touvron2023llama,
  title={Llama 2: Open foundation and fine-tuned chat models},
  author={Touvron, Hugo and Martin, Louis and Stone, Kevin and Albert, Peter and Almahairi, Amjad and Babaei, Yasmine and Bashlykov, Nikolay and Batra, Soumya and Bhargava, Prajjwal and Bhosale, Shruti and others},
  journal={arXiv preprint arXiv:2307.09288},
  year={2023}
}

\end{document}